\documentclass[11pt,a4paper]{article} 
\pdfoutput=1
\usepackage{jheppub}
\usepackage{enumitem}
\usepackage{epsfig} 
\usepackage{float} 
\usepackage[caption=false,labelformat=simple]{subfig}

\usepackage{wasysym} 
\usepackage{mathrsfs} 
\usepackage{graphicx} 
\usepackage{amsbsy} 
\usepackage{amscd} 
\usepackage{cancel}
\usepackage{multirow} 
\usepackage{tikz}
\usepackage{color}
\usepackage{xcolor}
\usepackage{slashed}
\usepackage{array}
\usepackage{tablefootnote}
\usepackage[normalem]{ulem}
\usepackage{color}
\usepackage{booktabs} 
\usepackage{amsmath,amssymb,amsfonts}%
\usepackage{amsthm}
\usepackage{textcomp}%
\usepackage{manyfoot}%
\usepackage{cleveref}

\usetikzlibrary{arrows,positioning,shapes.geometric} 
\usepackage[compat=1.1.0]{tikz-feynman}          
\usepackage[font=small,labelfont=bf]{caption}
\usepackage{diagbox}
\usepackage{multirow}
\usepackage{tabularx}
\usepackage{soul}  
\usepackage{listings} 
\usepackage{cleveref}
\usepackage{orcidlink}           
\usepackage{color}

\definecolor{somethingC}{rgb}{0.07, 0.04, 0.56} 

\usepackage{amsthm}

\newcommand{\stkout}[1]{\ifmmode\text{\sout{\ensuremath{#1}}}\else\sout{#1}\fi}

\newcommand{\dsum}{\displaystyle\sum}
\newcommand{\sumab}{\dsum_{a<b}}
\newcommand{\pta}{p_{T_a}}
\newcommand{\ptb}{p_{T_b}}
\newcommand{\dRab}{\Delta R_{ab}}

\title{Jet Substructure Probe on Scalar Leptoquark Models via Top Polarization}

\author[a]{Anupam Ghosh\,\orcidlink{0000-0003-4163-4491}\,}
\author[b]{\!\!, Partha Konar\,\orcidlink{0000-0001-8796-1688}\,}
\author[c]{\!\!, Tousik Samui\,\orcidlink{0000-0002-1485-6155}\,}
\author[d]{\!\!, Ritesh K. Singh\,\orcidlink{0000-0001-7838-6191}\,}

\affiliation[a]{Department of Physics, Indian Institute of Technology Guwahati, North Guwahati, 781039, Assam, India}
\affiliation[b]{Theoretical Physics Division, Physical Research Laboratory, Shree Pannalal Patel Marg, Ahmedabad, 380009, Gujarat, India}
\affiliation[c]{The Institute of Mathematical Sciences, IV Cross Road, CIT Campus, Taramani, Chennai, 600113, Tamil Nadu, India}
\affiliation[d]{Department of Physical Sciences, Indian Institute of Science Education and Research Kolkata, Mohanpur, 741246, Nadia, India}

\emailAdd{anupamg@rnd.iitg.ac.in}
\emailAdd{konar@prl.res.in}
\emailAdd{tousiks@imsc.res.in}
\emailAdd{ritesh.singh@iiserkol.ac.in}

\abstract{The study of leptoquarks and their couplings to fermions with different chiralities provides a powerful tool for distinguishing among different leptoquark models. As a case study, we focus on two specific third-generation scalar leptoquark models, $S_3$ and $R_2$, which differ in their electroweak quantum numbers and chiral structures of couplings to the top quark, leading to distinct top-quark polarization states. 
To enhance the efficacy of the analysis, we employ jet substructure techniques like Soft Drop, $N$-subjettiness, and our custom $b$-tagging method, along with other event variables. The analysis has been performed using both fixed radius and dynamic radius jet clustering algorithms. A multivariate analysis using a boosted decision tree (BDT) is performed to isolate signal from the Standard Model background. For a leptoquark mass of 1250 GeV, the analysis achieves a signal significance of up to $5.3\,\sigma$ at the 14 TeV HL-LHC. Furthermore, a $CL_s$-based profile likelihood estimator is applied to polarization-sensitive variables to discriminate between the two models. To enhance separation between the two models, an additional BDT classifier score is obtained by training a BDT network to distinguish between the $S_3$ and $R_2$ models. In the chosen signal region, the BDT classifier score provides a separation score of up to $3.2\,\sigma$, outperforming traditional variables such as $E_b/E_t$ and $\cos\theta_b$. 
}

\preprint{IMSc/2025/01}


\begin{document}
\maketitle

\newpage

\section{Introduction}
\label{intro}

Leptoquarks (LQs) are hypothetical particles that couple directly to quarks and leptons, arising naturally in various beyond the Standard Model (BSM) frameworks, including Grand Unified Theories, Pati-Salam unification, and models with extended flavor symmetries\,\cite{Pati:1973uk, Georgi:1974sy, Pati:1974yy, Langacker:1980js, Wudka:1985ef,deMedeirosVarzielas:2015yxm}. Scalar LQs represent a minimal and renormalizable extension of the Standard Model (SM) that can yield rich phenomenology at the TeV scale. While recent experimental updates have significantly reduced previous tensions in lepton flavor universality observables such as $R_{K*}$\,\cite{LHCb:2022zom, LHCb:2022qnv}, the broader flavor sector including longstanding anomalies in $R_{D*}$\,\cite{BaBar:2012obs,LHCb:2015gmp}, the muon $g-2$\,\cite{Muong-2:2006rrc, Muong-2:2021ojo}, and rare decays continues to motivate models with LQ mediators\,\cite{Ghosh:2022vpb, Ghosh:2023xbj,Crivellin:2025qsq}. Moreover, despite apparent flavor anomalies, scalar LQs remain compelling candidates for BSM physics due to their unique quantum numbers and testable collider signatures. Depending on the model realization, they can play a role in neutrino mass generation\,\cite{Cai:2017jrq, Zhang:2021dgl, Dorsner:2017wwn}, baryogenesis\,\cite{Gu:2011pf, Hati:2018cqp}, and dark matter scenarios\,\cite{Choi:2018stw, Mandal:2018czf, DEramo:2020sqv}. At the LHC, scalar LQs can be produced via QCD-mediated pair production or through Yukawa-driven single production, leading to characteristic final states with leptons and hadronic jets. Thus, direct searches at the LHC offer a powerful and complementary probe of LQ scenarios independent of low-energy anomalies, making them a continued focus of experimental interest.

Despite several scalar LQ models, we focus on third-generation scalar LQs, particularly the $SU(2)_L$ triplet $S_3$ and the doublet  $R_2$, which are the most theoretically motivated and phenomenologically rich extensions of the SM. Deviations in observables such as $R_{D*}$, the muon anomalous magnetic moment, and various B-physics channels suggest that new physics may preferentially couple to third-generation fermions. Both $S_3$ and $R_2$ can address these anomalies and remain important targets for direct searches at the LHC. Moreover, their preference for third-generation couplings is well motivated: it is less constrained by flavor precision data, aligns with observed hierarchies in the SM Yukawa sector, and is compatible with the principle of minimal flavor violation.

Although numerous phenomenological studies\,\cite{Gripaios:2010hv, Chandak:2019iwj, Bhaskar:2021gsy, Belanger:2021smw, Ghosh:2023ocz} have explored scalar LQ models, the CMS and ATLAS collaborations have conducted direct searches at the LHC -- probing various final states\,\cite{ATLAS:2020dsk, ATLAS:2024huc, ATLAS:2021oiz, CMS:2018lab, CMS:2024bnj, CMS:2020wzx, CMS:2019ybf, CMS:2018ncu}. The first and second-generation LQs can be searched in the final state of two oppositely charged leptons or a singly charged lepton with a significant amount of missing transverse momentum (MET) associated with jets. For example, a recent analysis by ATLAS\,\cite{ATLAS:2020dsk} excludes LQ of mass up to 1.8 TeV and 1.7 TeV at 13 TeV LHC with 139 $\rm fb^{-1}$ luminosity, considering branching ratio ${\rm BR(LQ}\to j e)=1$ and ${\rm BR(LQ}\to j \mu)=1$, respectively. Third-generation scalar LQs are also extensively studied in various final states by the ATLAS and CMS collaborations\,\cite{ATLAS:2023uox, CMS:2020wzx, ATLAS:2019qpq}. A recent ATLAS analysis\,\cite{ATLAS:2020dsf, ATLAS:2024huc} on scalar LQ pair production excludes third-generation scalar LQ masses up to 1240 GeV at $95\%$ CL, assuming a branching ratio ${\rm BR(LQ}\to t\,\nu)=1$.

In both the $S_3$ and $R_2$ models, scalar LQs exhibit couplings to the top quark and SM neutrino. If these LQs exist at the TeV scale, they can be copiously pair-produced at hadron colliders through their QCD coupling. Hence, the subsequent decay of each LQ into a top quark and a neutrino results in a unique collider signature observable at the LHC: two large radius (large-$R$) top jets, originating from the hadronic decay of boosted top quarks, and substantial MET. 
While kinematic variables constructed from the final state boosted top jets and MET enable us to separate LQ signals from the SM background, jet substructure-based polarization sensitive variables\,\cite{De:2020iwq, Dey:2021sug, De:2024puh} help pinpoint the specific model. For example, the top quark arising from the LQ described by the $S_3$ model tends to be left-chiral, whereas it is right-chiral in the $R_2$ model. This chirality structure is imprinted in the decay products of the top quark and manifests in the distribution of polarization-sensitive variables, such as the energy fraction $z_b = E_b/E_t$ of the $b$-quark and the angular variable $\cos\theta$ -- the angle between the $b$-quark momentum and the propagation direction of the top quark in the top rest frame. 
One can employ a multivariate analysis using a boosted decision tree (BDT) to improve model discrimination capability. This approach combines the characteristics of several variables as input for the BDT. The resulting BDT classifier score provides a powerful combined variable to distinguish one model from others. Furthermore, correctly identifying $b$-subjet within the boosted top jet is another crucial factor for constructing these variables, necessitating optimal use of jet substructure techniques. We utilize $N$-subjettiness to identify subjets within the boosted top jet and an IP-significance-based $b$-tagging method to pinpoint $b$-subjets.

In this work, our primary objective is to investigate the discovery prospects of scalar LQs in the $S_3$ and $R_2$ models at the 14 TeV HL-LHC. Following a potential discovery, we further examine the possibility of distinguishing between these two models using polarization-sensitive jet substructure observables. For this comprehensive analysis, we use multivariate analysis, in particular BDT, which, in principle, may be supplemented by machine learning (ML) techniques. The incorporation of ML approaches could yield additional improvements, such as using low-level jet features \cite{Guest:2016iqz,CMS:2020poo,ATLAS:2023gog,ATLAS:2023ixc,ATLAS:2014lvy}, enhanced tagging efficiencies \cite{CMS:2022prd,Kasieczka:2019dbj,ATLAS:2022qxm,Bols:2020bkb,Qu:2022mxj}, or effective discrimination between signal vs. background \cite{ATLAS:2023ixc,Bhattacharya:2020aid} or even between competing models. However, they often come at the expense of reduced interpretability, increased computational resources, and a lack of flexibility to adapt to changing strategies in similar analyses. Moreover, the systematics associated with ML-based analysis are still an open area of research. Since our central goal is to perform a study with existing intuitive polarization-sensitive methods, we confine ourselves to multivariate analysis, leaving the ML-based strategies for future exploration. To this end, we summarize the salient features of this work as follows: 
\begin{enumerate}[itemsep=0pt]
\item Conducted a detailed signal-background analysis of third-generation scalar LQ models using jet substructure observables and multivariate techniques with {\sc XGBoost}\,\cite{Chen:2016btl}.
\item Employed the dynamic radius jet algorithm\,\cite{Mukhopadhyaya:2023rsb} to enhance boosted signal detection and compared the efficacies of using the dynamic radius jet algorithm against the fixed radius. 
\item Implemented an IP-significance based\,\cite{CMS:2012feb, ATLAS:2015thz, CMS:2017wtu, ATLAS:2018sgt, ATLAS:2019bwq, ATLAS:2022qxm} $b$-subjet tagging method within boosted top jets. 
\item Utilized polarization-sensitive variables, such as $b$-subjet energy fraction, angular variable, and a BDT trained combined variable for model discrimination; the methodology is broadly applicable to other SM/BSM scenarios with boosted final states.
\end{enumerate}

The rest of the manuscript is organized as follows: 
Section \ref{sec:model}  discusses the underlying theoretical framework of the scalar LQ models. We also provide a brief overview of production processes and existing collider constraints. 
Since investigating QCD jets is an integral part of the present analysis, various jet algorithms and jet substructure methods used in this study are discussed in Section \ref{jet_algo}. We also touched upon different jet-specific issues, such as the jet algorithm considering the dynamic radius and the possibility of tagging a $b$-subjet inside the top-like jet.
In the following Section \ref{Background section}, we delve into the collider methodology, explaining the targeted scalar LQ signal topologies and detailed background processes, specifying suitable benchmark points for our analysis. We detail the primary event selection process based on these simulated signal and background events, followed by constructing appropriate high-level variables and pointing out some of the effective variables not only to suppress the background events but also those that are efficient in distinguishing between the $S_3$ and $R_2$ models. Subsequently, we employ multivariate analysis techniques on these input variables to calculate the corresponding signal efficiencies.
Since our goal is beyond the discovery of LQ signals and stepping into discriminating one signal model from another, we advocate a log-likelihood ratio test with different polarization-sensitive variables in Section \ref{sec:distinguigh-models}.
Finally, the outcome of the findings has been discussed in Section~\ref{sec:results} before summarizing our key findings in Section~\ref{sec:conclsn}. 

\section{Phenomenological Framework}
\label{sec:model}
Leptoquarks are bosons and have fractional electric charges. They carry charges under both $SU(3)$ and $SU(2)$ groups, and have simultaneous coupling to quark and lepton. Leptoquarks carry baryon and lepton numbers. If LQs exhibit interactions that violate either baryon or lepton number conservation, they must be heavy, typically on the order of $10^{16}$~GeV -- due to the stringent limits from the lack of observed proton decay to date and large Majorana neutrino masses. Alternatively, we consider the Buchm{\"u}ller-R{\"u}ckl-Wyler (BRW) framework\,\cite{Buchmuller:1986zs}, which assumes lepton and baryon number-conserving interaction. In this framework, LQs need to satisfy much weaker bonds and can have masses $\mathcal{O}(100)$ GeV. It is also possible to obtain the flavor diagonal couplings, which means LQ couples with the same generation of the SM quark and lepton pair. The same-generation interactions avoid the possibility of flavor-changing neutral currents (FCNC)\,\cite{Mitsou:2004hm}.


\begin{table}[b!]
	\begin{center}
		\resizebox{1.0\columnwidth}{!}{%
		    \begin{tabular}{@{}lllll@{}}
				\toprule
			LQ &  \scriptsize$\{ SU(3), SU(2)_L,$  & \scriptsize $SU(2)_L$ & Yukawa interaction & Decay modes   \\
			model & \scriptsize$U(1)_Y \}$	& structure & (flavor basis) & (third-gen.) \\  
				\midrule
				\multirow{ 2}{*}{$S_3$} & \multirow{ 2}{*}{({\bf $\bar{3}, 3, \frac{1}{3}$}) } &\multirow{ 2}{*}{ $\scriptsize  \begin{pmatrix}
					S_3^{4/3} \\  
					S_3^{1/3}\\
					S_3^{-2/3}
				\end{pmatrix} $ }&  \multirow{ 2}{*}{ $y_{LL}^{ij} \bar{Q}_i^C ~i\tau_2~(\tau_k S_3^k)~ L_j + h.c. $ }  & $S_3^{\frac{4}{3}}(b, \tau)$, $S_3^{\frac{1}{3}}((t, \tau), (b, \nu_\tau))$,   \\
			 &  &  &   & $S_3^{-\frac{2}{3}}(t, \nu_\tau)$\\
 \hline
				\multirow{ 2}{*}{$R_2$} & \multirow{ 2}{*}{({\bf $3, 2, \frac{7}{6}$ }) } &\multirow{ 2}{*}{ $\scriptsize  \begin{pmatrix}
						R_2^{5/3} \\  
						R_2^{2/3}
					\end{pmatrix} $ }&  $(y_{LR}^{ij}~ \bar{Q}_i~ l_{R,j}~ R_2 $   & $R_2^{\frac{5}{3}}(t , \tau)$,   \\
				&  &  & $-  y_{RL}^{ij}~ \bar{u}_R^i~ R_2 ~ i\tau_2~ L_j) + h.c.$   & $R_2^{\frac{2}{3}} ((t , \nu_\tau), (b, \tau))$\\
 \hline
 				\multirow{ 2}{*}{$\tilde{R}_2$} & \multirow{ 2}{*}{({\bf $3, 2, \frac{1}{6}$ }) } &\multirow{ 2}{*}{ $\scriptsize  \begin{pmatrix}
 		\tilde{R}_2^{\frac{2}{3}} \\  
 		\tilde{R}_2^{-\frac{1}{3}}
 	\end{pmatrix} $ }&  $(\tilde{y}_{\overline{LR}}^{ij}~ \bar{Q}_i~\tilde{R}_2~ \tilde{\nu}_{R,j}$   & $\tilde{R}_2^{\frac{2}{3}}((t , \tilde{\nu}_R),(b , \tau))$,   \\
 &  &  & $- \tilde{y}_{RL}^{ij}~ \bar{d}_R^i~ \tilde{R}_2 ~ i\tau_2~ L_j) + h.c.$   & $\tilde{R}_2^{-\frac{1}{3}}((b , \nu_\tau),(b ,  \tilde{\nu}_R))$\\
 \hline
	\multirow{ 2}{*}{$S_1$} & \multirow{ 2}{*}{({\bf $\bar{3}, 1, \frac{1}{3}$ }) } &\multirow{ 2}{*}{ $\scriptsize  \begin{pmatrix}
		S_1^{1/3} 
	\end{pmatrix} $ }&  $(y_{LL}^{ij}~ \bar{Q}_i^C~ S_1~i\tau_2~L_{j}$   & \multirow{ 2}{*}{$S_1^{\frac{1}{3}}((t, \tau), (b, \nu_\tau), (b, \tilde{\nu}_R))$ }  \\
&  &  & $+y_{RR}^{ij}~ \bar{u}_R^{C,i}~ S_1 ~ l_{R,j} + y_{\overline{RR}}^{ij}~ \bar{d}_R^{C,i}~ S_1 ~ \tilde{\nu}_{R,j}) + h.c.$   & \\
 \hline
$\tilde{S}_1$ & ({\bf $\bar{3}, 1, \frac{4}{3}$ }) & $\scriptsize  \begin{pmatrix}
 		\tilde{S}^{4/3}_1 \end{pmatrix} $ &  $(y_{RR}^{ij}~ \bar{d}_R^{C,i}~ \tilde{S}_1~l_{R,j}) + h.c.$   & $\tilde{S}_1^{\frac{4}{3}}(b, \tau)$ \\
 \hline
$\bar{S}_1$ & ({\bf $\bar{3}, 1, -\frac{2}{3}$ }) & $\scriptsize  \begin{pmatrix}
	\bar{S}^{-\frac{2}{3}}_1 \end{pmatrix} $ &  $(y_{\overline{RR}}^{ij}~ \bar{u}_R^{C,i}~ \bar{S}_1~\tilde{\nu}_{R,j}) + h.c.$   & $\bar{S}_1^{-\frac{2}{3}}(t, \tilde{\nu}_R)$ \\
	\bottomrule
    \end{tabular}
    } 
		\caption{Scalar LQ models, including their gauge-invariant interaction terms and charges under the SM gauge groups, are presented. The third-generation scalar LQ decay modes are presented in the last column.}
		\label{tab:LQ}
	\end{center}
\end{table}
\begin{figure}[htbp!]
	\begin{center}
		\includegraphics[angle=0,scale=0.8]{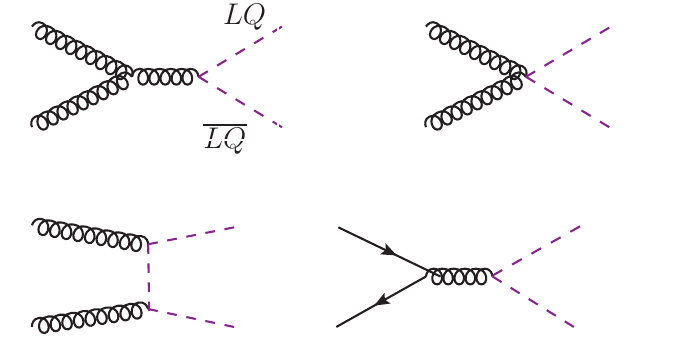}\vspace{1mm}
		\caption{Leptoquark pair production processes at leading order, coming from Eq.~(\ref{eq:Lkin}).} 
		\label{fig:feynman_dia}
	\end{center}
\end{figure}
All possible scalar LQ models, their gauge-invariant interaction with the SM fermions, and their charges under the SM gauge groups are listed in Table \ref{tab:LQ}. The LQ model conventions follow those in Ref.\,\cite{Dorsner:2016wpm}. Here, $C$ and $h.c.$ are the complex and hermitian conjugation,
respectively, and $\tau_i~(i=1,2,3)$ are the Pauli matrices. The Yukawa coupling  $y_{LR}^{ij}$ corresponds to LQ coupling with a SM left-handed quark doublet of $i$-th generation with the right-handed lepton ($j$-th generation) and so on. The bar in the coupling written for LQ couples with the BSM right-chiral neutrinos ($\tilde{\nu}_R$). We are interested in studying third-generation LQ, which means that an LQ will decay into a third-generation SM quark after production. The corresponding decay modes for same-generation coupling are presented in the last column of Table \ref{tab:LQ}.

In proton-proton collisions, LQs can be pair-produced and singly produced. For scalar LQ, a single coupling between the LQ and quark-lepton pair determines the coupling strength of the Yukawa interaction. On the other hand, vector LQs require two additional coupling constants to account for interactions associated with the electric quadrupole moment and magnetic moment\,\cite{Hewett:1997ce}. Therefore, vector LQ production involves these additional parameters, and their complete interpretation lies beyond the scope of this paper. 

In this study, we focus on the pair production of scalar LQs. The kinetic Lagrangian for a generic scalar LQ ($S$) is given by,
\begin{equation}
{\cal{L}}_{\rm{kin}} =(D_\mu S)^\dagger (D^\mu S)  - M_S^2 S^\dagger S \,. 
\label{eq:Lkin}
\end{equation}
where covariant derivative is given as,
\begin{equation}
	D_\mu = \partial_\mu -i g_s \dfrac{\lambda^a}{2} G^a_\mu -i g\dfrac{\sigma^i}{2} W^i_\mu -i g^{'} Y B_\mu \,.
\end{equation}
Here, $g_s$ is the strong coupling, $g$ and $g'$ are the electroweak couplings, and $Y$ is the hypercharge. In a proton-proton collider, LQ (along with its anti-particle) production is dominated by strong production processes through quark-antiquark and gluon annihilation diagrams. LQ is also likely produced through weak boson exchange, but that contribution is tiny compared to the strong production. The leading-order (LO) production processes, illustrated in Figure \ref{fig:feynman_dia}, are identical for all six scalar LQ models. Note that, pair production of third-generation scalar LQ is driven solely by the strong coupling constant and the LQ mass, rendering it independent of BSM couplings\footnote{Third-generation LQ can also be pair produced through $b\bar{b}$-initiated $t$-channel diagrams mediated by lepton exchange if the corresponding LQ has coupling with $b$-quark (like $R_2^{2/3}$ but unlike $S_3^{-2/3}$)\,\cite{Mandal:2015lca}. However, this process is significantly suppressed due to the small parton distribution functions (PDFs) of the initial-state $b$ quarks.}. In contrast, single production is model-dependent and sensitive to the LQ-quark-lepton couplings. Next-to-leading order (NLO) QCD corrections for scalar LQ pair production, including matching to the {\sc Pythia8} parton shower, have been investigated in Refs.\,\cite{Mandal:2015lca, Ghosh:2023ocz, Kramer:2004df}.

Our study considers only those LQ models that do not couple with any BSM right-handed neutrino. After pair production, we are interested in the final state of $t\bar{t}$ associated with significant MET~\footnote{Many dark matter models, such as WIMPs or multicomponent scenarios involving WIMPs and axions, predict final states with a top quark pair and significant MET. We refer the reader to Refs.\cite{Ghosh:2022rta, Ghosh:2023xhs, Ghosh:2024nkj, Ghosh:2024boo, Ghosh:2021noq} for a comprehensive study of the dark sector phenomenology and detailed collider analyses in the large-$R$ jets plus MET channel at the LHC, including jet substructure variables and angular observables.}, and hence, we are focusing only the $S_3$ and $R_2$ models. We want to explore the 2/3-e electromagnetic-charged scalar LQs from the $S_3$ and $R_2$ models. The Yukawa interactions are given below, which are obtained by expanding the $SU(2)_L$ triplet and doublet of $S_3$ and $R_2$, as in Table \ref{tab:LQ}.
\begin{align}
	{{\cal{L}}_{\rm{Int}}^{S_3^{\tiny \frac{2}{3}}}} &= y_{{LL}}\  \overline{t_L^C}\,\nu_{\tau}\,S_3^{-\frac{2}{3}} + h.c., \label{eq:LInt} \\
	{{\cal{L}}_{\rm{Int}}^{R_2^{\tiny \frac{2}{3}}}} &= y_{{RL}} \ \overline{t_R}\,\nu_{\tau}\, R_2^{\frac{2}{3}}   +  y_{{LR}}\ \overline{b_L}\,\tau_R\,R_2^{\frac{2}{3}} + h.c. \label{eq:LInt2}
\end{align}
The rate of the pair production processes $pp \to S_3^{2/3} S_3^{-2/3}$ and $pp \to R_2^{2/3} R_2^{-2/3}$ are the same at the LHC. Moreover, $R_2^{2/3}$ has two decay modes controlled by the respective Yukawa couplings. However, if we assume $R_2^{2/3}$ to decay entirely into the top quark and a neutrino, the final production cross section into the $t\bar t$ plus MET final state remains equal for both processes. 

Leptoquark masses below 1 TeV have already been excluded by CMS and ATLAS searches\,\cite{CMS:2020wzx, ATLAS:2020dsf, ATLAS:2024huc}. Consequently, a TeV-scale LQ would decay to produce a boosted top quark, whose hadronic decay results in a three-prong boosted top jet. Therefore, we focus on a final state comprising two boosted top jets accompanied by significant MET as a promising signature for probing these compelling BSM particles. The presence of two neutrinos in the final state from the LQ decay is the source of the large MET. In the following few subsections, we detail the collider analysis of this model using sophisticated multivariate analysis (MVA) using a BDT. Furthermore, we compare our results obtained through the fixed and dynamic radius jet algorithms. 

After discussing the discovery prospects, Section \ref{sec:distinguigh-models} further focuses on distinguishing between two models of scalar LQs, $S_3^{2/3}$ and $R_2^{2/3}$. These LQs have the same electromagnetic charges, have identical production cross sections, and produce the same final states at the LHC. However, as shown in Eqs.~\ref{eq:LInt} and \ref{eq:LInt2}, the top quarks produced by $S_3^{2/3}$ and $R_2^{2/3}$ exhibit opposite chiralities. By analyzing various kinematic and angular variables sensitive to top quark polarization, we explore the prospects for distinguishing these two models at the LHC~\footnote{Interesting to note that the distribution of the $i$-th decay particle, such as the $b$ quark from a right-handed top quark, is identical to the distribution of the $\bar{b}$ anti-quark originating from the decay of a left-handed antitop quark.}.

In recent analysis\,\cite{Ghosh:2023ocz}, authors attempted to distinguish the same LQs by tagging the $b$-subjet within the boosted top jet using the separation variable $\Delta R(b, J_t)$, where it was found that rather poor discrimination of maximum one standard deviation ($1 \sigma$) could be achieved at the High-Luminosity LHC (HL-LHC), utilizing the observable, $z_b=E_b/E_t$, where $E_b$ and $E_t$ are the energies of the $b$-tagged subjet and the boosted top jet, respectively. Further, it was observed that some of the events obtained unphysical $z_b>1$, due to the contamination from poor reconstruction.

In this work, we analyze the constituents of the boosted top jet to tag a $b$-subjet within it (detailed procedure described in Subsection \ref{IP_algo}). We examine discrimination prospects using the different variables $z_b$, $\cos\theta_b$ ($\theta_b$ being the angle of the $b$-jet in the rest frame of the reconstructed boosted top jet), and the combination of them with other observables. Our results show that utilizing the combined observable achieves discrimination beyond three standard deviations ($3 \sigma$) at the HL-LHC. This analysis is performed using both fixed radius and dynamic radius jet algorithms. The scope of the prescribed methodology for model discrimination is equally relevant to other BSM scenarios with similar final states.

\section{Jets and Substructure }
\label{jet_algo}
In this work, we employ a set of jet substructure techniques to tag the top-like jet more effectively, within which a $b$-subjet is expected to be present. For our analysis, we first utilize two jet algorithms, namely the anti-$k_t$\,\cite{Cacciari:2008gp} and the dynamic radius anti-$k_t$\,\cite{Mukhopadhyaya:2023rsb} algorithms, for jet reconstruction. We then apply the Soft Drop\,\cite{Larkoski:2014wba} grooming procedure, followed
by the $N$-subjettiness\,\cite{Thaler:2010tr, Thaler:2011gf} technique to find the subjets within the large-$R$ jet. Subsequently, we implement IP-significance-based $b$-subjet tagging within the {\sc Delphes} framework\,\cite{deFavereau:2013fsa}, following the standard $b$-tagging procedure. In the subsequent subsections, we briefly describe and discuss these methods for completeness.

\subsection{Fixed and Dynamic Radius Jet Algorithm}
\label{DR_algo}
Our study involves jets arising from boosted top quarks. The showers from the decay products of the boosted top quark form a jet, which is characteristically wider than the QCD jets originating from (anti-)quarks or gluons. Jet substructure techniques have been proven to be powerful in improving the tagging of top quarks from this top-initiated jet. Before analyzing the jets at the substructure level, one requires a prescription to cluster different hadronic deposits in the form of jets from an event. The $k_t$-type sequential recombination jet clustering algorithms are commonly used in the context of LHC since these algorithms are infrared and collinear safe. These algorithms take a list of four momenta in an event as input, then calculate pairwise distance $d_{ij}$ for the pair indexed by $i$, $j$, and beam distance $d_{iB}$ for the $i$-th entry:
\begin{eqnarray}
&& d_{ij} = \min\left({p_T}_i^{2n}, {p_T}_j^{2n}\right) \Delta R^2_{ij}, {\ \rm and} \\
&& d_{iB} = {p_T}_i^{2n} R^2,
\end{eqnarray}
where $\Delta R_{ij}$ is the Euclidean distance between $i$-th and $j$-th four momenta in the pseudorapidity-azimuth ($\eta-\phi$) plane. In the popular choices, the exponent $n$ takes values $-1$, $0$, or $1$ corresponding to anti-$k_t$ (AK), Cambridge-Aachen (CA), or $k_t$ (KT) algorithm, respectively. The radius parameter $R$ is an input for the typical radius of the jet.
For the large-$R$ jet study, the typical choices range from 0.8 to 1.5, while the same varies between 0.4 and 0.5 for a light jet produced from a quark or gluon. However, one can set only one number in a particular event. Therefore, one can only get either narrow jets or jets with large radius in a single run of the algorithm. If one expects varying-sized jets in an event, they need to run the algorithm twice, then perform an overlap removal to find narrow and large-$R$ jets exclusively. In our signal, the radius varies because of the level of boost in each decaying top quark, which leads to the formation of a jet. Several attempts have been made to find varying-sized jets in a single run of the jet algorithm. Some of such attempts are variable radius jets\,\cite{Krohn:2009zg}, fuzzy jets\,\cite{Mackey:2015hwa}, dynamic radius jets\,\cite{Mukhopadhyaya:2023rsb}, PAIReD jets\,\cite{Mondal:2023law}, Jet-SIFTing\,\cite{Larkoski:2023nye}, SHAPER\,\cite{Ba:2023hix}.  
In this work, we have used the dynamic radius jet algorithm\,\cite{Mukhopadhyaya:2023rsb} in addition to the traditional fixed radius algorithms to compare the efficacy of the former with respect to the latter.

In the dynamic radius algorithm, the fixed radius $R$ is replaced by $R_0 + \sigma_i$ for the $i$-th pseudojet during the evolution of the jets through the merger of four-momentum. The additive radius modifier $\sigma_i$ is essentially the $p_T$-weighted standard deviation of the inter-constituent distances in the $\eta-\phi$ plane, defined as
\begin{equation}
	\sigma_i^2 = \dfrac{\sumab \pta\,\ptb\ \dRab^2 }
	{\sumab \pta\,\ptb } - \left(\dfrac{\sumab \pta\,\ptb\ \dRab}
	{\sumab \pta\,\ptb }\right)^2, \label{eqn:var}
\end{equation}
where the indices $a$ and $b$ run over the constituents inside the $i^{\rm th}$ evolving pseudojet. 
The present algorithm dynamically modifies the radius parameter of each evolving jet as it captures more and more constituents, beginning at an initial radius $R_0$. The $\sigma_i$ is related to the energy correlation functions\,\cite{Larkoski:2013eya, ATLAS:2019kwg} usually used to study jet substructure, and its value is expected to be greater for a boosted large-$R$ jet and smaller for narrow jets. Thus, it modifies the radius parameter appropriately according to the size of the jet. The utility and other implications have been detailed in Ref.\,\cite{Mukhopadhyaya:2023rsb}. As in the fixed radius algorithm, the exponent $n$ can take values $-1$, 0, or 1 corresponding to dynamic radius anti-$k_t$ (DR-AK), dynamic radius Cambridge-Aachen (DR-CA), or dynamic radius $k_t$ (DR-KT) algorithm.

This dynamic radius algorithm is particularly useful in rejecting QCD backgrounds in a top jet study since it helps set a narrower radius for the QCD jets while allowing for a larger radius for boosted large-$R$ jets. This work primarily focuses on boosted top quarks, which results in a large-$R$ jet. Therefore, the dynamic radius algorithm is helpful in rejecting the vast QCD background and thus helping in selecting signals. Typical top-like boosted jet is characterized by a fixed radius $R$  between 0.8 and 1.5. For the dynamic radius algorithm, our initial radius $R_0$ choice ranges from 0.6 to 1.2. For both fixed and dynamic radius, we classify the jets, clustered with $R$ and $R_0$, respectively, in the ranges mentioned, `large-$R$' jets. 


\subsection{Soft Drop Grooming}
\label{Soft-drop}
In the analysis of top jets, the jet substructure study plays an important role in tagging a jet as a top jet. The substructure analysis involves the constituents of the jet. On the other hand, in the particle colliders such as the LHC, unwanted contaminations such as pileup, underlying events, and soft radiations affect the jet substructure variables, contributing to a poor resolution of these variables. Therefore, it is essential to remove these contaminations before constructing and analyzing jet substructure observables. There are quite a few grooming methods, {\it e.g.}, Modified Mass-Drop Tagger (mMDT)\,\cite{Butterworth:2008iy, Dasgupta:2013ihk}, Trimming\,\cite{Krohn:2009th}, Pruning\,\cite{Ellis:2009su, Ellis:2009me}, Dynamical Grooming\,\cite{Mehtar-Tani:2019rrk, Mehtar-Tani:2020oux}, etc. These methods help in mitigating contamination inside jets while preserving the core structure of a jet. In this work, we have used Soft Drop declustering\,\cite{Larkoski:2014wba}, a widely used grooming method, to remove soft and wide-angle radiations from a jet. Soft Drop improves jet mass resolution by grooming away the soft and wide-angle radiation, which helps in enhancing discrimination between signal and background. 

The Soft Drop algorithm is applied to jets after reclustering with a sequential algorithm (e.g., Cambridge-Aachen, $k_t$, or anti-$k_t$). The algorithm has the following steps:
\begin{enumerate}[itemsep=0pt]
\item Take a jet and recluster is using a sequential clustering algorithm.
\item Iteratively decluster the jet into two subjets: $i$ and $j$.
\item For each pair of subjets, check the Soft Drop condition:
	\[
	 \frac{\min({p_T}_i, {p_T}_j)}{{p_T}_i + {p_T}_j} > z_{\text{cut}} \left( \frac{\Delta R_{ij}}{R_{\rm SD}} \right)^{\beta_{\rm SD}},
	\]
	where $p_{T,i}$ and $p_{T,j}$ are the transverse momenta of the subjets, $\Delta R_{ij}$ is distance between them.

\item If the Soft Drop condition is satisfied, the grooming process stops, and the jet is considered ``groomed".
\item If the condition is not satisfied, the softer subjet is discarded, and the process continues with the harder subjet from step 2.
\end{enumerate}
The Soft Drop condition has two parameters (1) $z_{\text{cut}}$: energy fraction threshold and (2) $\beta_{\rm SD}$: angular exponent controlling the dependence on the subjet separation.
In the special case $\beta_{\rm SD} = 0$, the method reduces to the mMDT. In this case, the grooming depends only on the momentum fraction $z_{\rm cut}$ and not on the angular separation.
However, for larger values of $\beta_{\rm SD}$, it returns almost ungroomed jets, while smaller values grooms the more strongly. For this work, we have used $\beta_{\rm SD} = 1.0$ and $z_{\rm cut} = 0.1$. 
The radius $R_{\rm SD}$ is the characteristic jet radius used during the clustering. For the fixed radius algorithm, we have used $R_{\rm SD} = R$, and for the dynamic radius algorithm $R_{\rm SD}$ is taken to be the final acquired jet radius $R_{\rm fin} \equiv R_0+\sigma_i$.

\subsection{Subjets using $N$-subjettiness}
\label{N-subjettiness}
After the grooming of the large-$R$ jets via Soft Drop, we use another jet substructure method, namely the $N$-subjettiness\,\cite{Thaler:2010tr}, for finding the three subjets expected inside the boosted top jet. Therefore, we are particularly interested in 3-subjettiness. The $N$-subjettiness method is based on finding the subjet by fitting $N$ number of subjet axes to the $N$ possible lobes or ``prongs'' inside the full jet. This is performed by optimizing the $N$-subjettiness measure\,\cite{Thaler:2011gf}
\begin{eqnarray}
\tilde{\tau}_N^{(\beta)} = \frac{1}{d_0}\sum_{i\in {\rm jet}} {p_T}_i \min\left\{ \Delta R_{i, a_1}^\beta, \Delta R_{i, a_2}^\beta, \cdots, \Delta R_{i, a_N}^\beta \right\}, 
\label{eqn:nsubjettiness}
\end{eqnarray}
where $i$ runs over the constituents of the jet, and $a_k$ represents the four momentum of the $k$-th axis. The prefactor $1/d_0$ is a normalization factor. The exponent $\beta$ is a parameter of choice, which keeps the $N$-subjettiness measure general like the jet substructure variable generalized angularities\,\cite{Berger:2003iw,Almeida:2008yp,Ellis:2010rwa,Larkoski:2014pca,Gras:2017jty,Nayak:2019quy} or energy correlation functions\,\cite{Larkoski:2013eya}. We have used $N$-subjettiness following the method suggested in Ref.\,\cite{Thaler:2011gf}, which is implemented in {\sc Fastjet Contrib}\,\cite{Thaler:2010tr}. The minimization is performed over the set of axes \{$a_1$, $a_2$, $\cdots$, $a_N$\} to achieve the final 
\begin{eqnarray}
	\tau_N^{(\beta)} = \min_{\{a_1, a_2, \cdots, a_N\}} \tilde{\tau}_N^{(\beta)}.
\end{eqnarray}

In this method, the optimization is performed using a variation of Lloyd's $k$-means clustering algorithm for finding $k$ numbers of clusters\,\cite{Lloyd:1982zni}. We briefly discuss the essential steps of the algorithm below:
\begin{description}[itemsep=0pt,topsep=2pt]
\item[Initialization:] The initial seed axes for further optimization are found in this step. References\,\cite{Thaler:2011gf,Stewart:2015waa} discuss a number of choices for finding the seed axes.

\item[Iteration:] After the initialization, the following iterative steps are performed until a desired accuracy is reached. 
\begin{description}[topsep=0pt]
\item[Assignment:] After setting trial axes in an iteration, each jet constituent is assigned to the axis that is closest (as per distance measure $\Delta R$) to it. Thus a subcluster containing one or more constituents of the jet is assigned to each of subjet axes. In geometric sense, this step divides the whole jet area into sub areas corresponding to each subjet axis.

\item[Update:] This iterative step updates the old axes to the new ones in order to minimize the $N$-subjettiness measure given in Eq.~\ref{eqn:nsubjettiness}. In this work, the update is performed following Ref.\,\cite{Thaler:2011gf} for general values of $1\leq \beta < 3$. 

\item[Convergence Check:] Iteration over the last two steps, {\it viz.} Assignment and Update, is performed until a desired accuracy is achieved. The average shift of axes in the $\eta - \phi$ plane is usually defined as the accuracy measure. For $(r + 1)^{\rm th}$ iteration, the accuracy is
defined as
\begin{equation}
\Delta^{(r+1)} \equiv \frac{1}{N}\sum_{k=1}^{N} \Delta R_{a_k^{(r)},a_k^{(r+1)}}, 
\end{equation}
where the sum $k$ runs over subjet axes. 
\end{description}

\end{description}
In the above procedure, there are two major places for choices (1) seed axes and (2) the exponent $\beta$. We have used {\sc WTA KT} axes as the seed axes for the minimization procedure and $\beta = 1.0$. Initially the optimal $N$-subjettiness measure $\tau_N^{(\beta)}$ suggested observable for $N$ pronged jets. However, it has been realized that the $N$-subjettiness ratios $\tau_N^{(\beta)}/\tau_{N-1}^{(\beta)}$ is more effective for $N$ pronged jet\,\cite{Thaler:2010tr,Thaler:2011gf}. In our case, we are interested in the top jet, where 3 subjets are expected. So, we have used $\tau_3^{(\beta)}/\tau_2^{(\beta)}$.

\subsection{Tagging $b$-subjet inside Boosted Top Jet}
\label{IP_algo}
Our signal topology consists of (anti-)top quark, which decays to a (anti-)bottom and a $W$ boson. In the hadronic decays of $W$, the top quark gives rise to a large-$R$ jet containing three subjets inside it. In the boosted top jet, therefore, one $b$-subjet is expected. Thus, $b$-tagging helps enhance signal efficiency and reduce QCD background. Additionally, since one of our aims is to construct polarization-sensitive variables, tagging the $b$-subjet inside the boosted top jet is a crucial step forward to enhance the sensitivities of such studies. The necessity of this tagging arises because polarization-sensitive observables, such as angular distributions and energy fraction variables, are directly influenced by the spin correlation between the parent top quark and its decay products. Tagging of the $b$-subjet allows for the reconstruction of these observables, thereby improving the discrimination power between different polarization states.

The preliminary works of $b$-tagging were primarily based on impact parameter (IP) based algorithms\,\cite{CMS:2012feb, ATLAS:2015thz, CMS:2017wtu, ATLAS:2018sgt, ATLAS:2019bwq, ATLAS:2022qxm}. In the IP-based algorithms, the tracks are first listed inside a given jet. For each track, IP significance, defined by the ratio of the IP of the track to the uncertainty in the measurement of the IP, is calculated. A sign is assigned to the IP significance depending on whether the IP forms an angle less than ($+$ sign) or greater than ($-$ sign) $\pi/2$. For each jet, the tracks are then sorted in ascending order according to the signed IP significance. There are two variations of the IP significance based on whether a 2-dimensional IP (IP2D) is taken or a 3-dimensional IP (IP3D) is taken. Here, IP2D is basically the two-dimensional component in the $x$-$y$-plane (without the $z$ component) of the IP, while 3D includes the $z$ component. One then chooses either the second or third track's signed IP significance to discriminate $b$-jet from other jets. 

All these algorithms are primarily used for narrow jets and not for the subjets. One might easily extend the $b$-jet tagging prescription to the $b$-subjet tagging and expect similar efficiency. In our work, we did not really perform a detailed analysis of the $b$-subjet tagging; instead, we have taken a similar approach as implemented in the {\sc Delphes}\,\cite{deFavereau:2013fsa} $b$-tagging module. In this approach, the jets are matched with the original parton flavors of the hard process. This means that a jet is temporarily assigned the flavor of the parton closest to it within a cut-off distance $R_{\rm cut}$ in the $\eta-\phi$ plane. Then $b$-tagging efficiency and rates for other jets $b$-jet (mistagging rate) are taken from an analysis. Then the efficiency or mistagging rate is applied probabilistically based on the assigned jet flavors. For our analysis, subjet transverse momentum (measured in units of GeV) dependent efficiency and mistagging rates are given in Table~\ref{tab:effi} \cite{CMS:2012feb}.

\begin{table}[H]
\begin{center}	\begin{tabular}{cl}
\hline
Originating parton & Efficiency to be tagged as $b$-subjet\\
\hline
$b$-quark &		$21.25\tanh(0.0025 p_T)/(1+0.063p_T)$ \\
$c$-quark & $0.25\tanh(0.018 p_T)/(1+ 0.0013 p_T)$ \\
light quarks and gluon & $0.01+0.000038 p_T$\\
\hline
	\end{tabular}
\caption{Subjet $b$-tagging efficiency and mistag rates as a function of subjet transverse momentum ($p_T$, in GeV).}
\label{tab:effi}
\end{center}
\end{table}

We have implemented this tagging and mistagging efficiency for the subjets of a large-$R$ jet within the {\sc Delphes} framework, used for detector simulation in this work, as a separate module. The value for $R_{\rm cut}$ has been taken to be 0.2. In this method, multiple subjets within a large-$R$ jet may be tagged as $b$-subjets if they lie within $R_{\rm cut}$. In such situations, we choose the one with the highest IP3D significance for the second track. We further note that a more realistic and efficient $b$-subjet tagging can be performed using MVA or ML methods. However, those are beyond the scope of this work, and we leave them for future study.

\begin{table}[!h]
	\begin{center}
			\begin{tabular}{|c|c|c|c|}
				\hline
				\multirow{2}{*}{\bf Method} & \multirow{2}{*}{\bf Parameter} & \multicolumn{2}{c|}{\bf Parameter Values} \\
				\cline{3-4}
				 & & \bf for $R_2$ & \bf for $S_3$\\
				\hline
				anti-$k_t$ & $R$ & 1.2 & 0.8\\
				\hline
				dynamic-radius & $R_0$ & 0.8 & 0.6\\
				\hline
				\multirow{2}{*}{soft drop} & $\beta_{SD}$ & 1.0 & 1.0\\
				\cline{2-4}
				& $z_{\text{cut}}$ & 0.1 & 0.1\\
				\hline
				\multirow{2}{*}{$N$-subjettiness} & seed axes & {\sc WTA KT} & {\sc WTA KT}\\
				\cline{2-4}
				 & $\beta$ & 1.0 & 1.0 \\
				 \hline
				 \multirow{2}{*}{$b$-tagging} & track number (for IP3D) & 2nd & 2nd \\
				  \cline{2-4} & $R_{\rm cut}$ & 0.2 & 0.2 \\
				\hline\hline
			\end{tabular} 
		\caption{Jets and jet substructure methods and corresponding parameter values used in our analysis for large radius jets.}
		\label{tab:jet-params}
	\end{center}
\end{table}

In Table \ref{tab:jet-params}, we list all the parameters used in the jets and jet substructure methods described in our analysis. The radius parameters for the anti-$k_t$ and dynamic radius anti-$k_t$ have been optimized for the signal $R_2$ and $S_3$ separately, and it will be discussed further in sections below. 

\section{Signal and Background Processes}
\label{Background section}
\begin{figure}[tb!]
	\centering
	\includegraphics[scale=0.62]{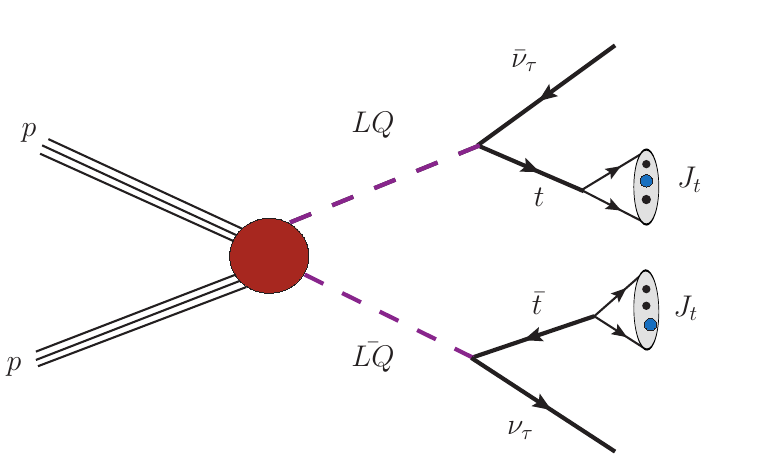}
	\caption{
		A schematic diagram of two top-like jets and MET resulting from the pair production of scalar LQs at the LHC is presented.
	}
	\label{sig_FJ}
\end{figure}

We study the pair production of LQs, where each LQ subsequently decays into a top quark and a SM neutrino. The top quark appears from the decay of a TeV-scale LQ and is expected to have a significant boost from such production. Consequently, the hadronic decay of the top quark often manifests as a three-pronged jet. The signal topology is as follows:
\begin{eqnarray}
	pp\to S \bar{S} \to (t \nu_\tau) (\bar{t} \nu_\tau) \equiv 2J_t + {\rm MET}~,
	\label{sig_topo}
\end{eqnarray}
where $S$ is either of the scalar LQs $S_3^{+ 2/3}$ or $R_2^{+ 2/3}$. A diagrammatic representation of the signal topology is shown in Figure \ref{sig_FJ}. The two neutrinos in the final state remain invisible to the detector. They are the source of a large MET derived from the total imbalance of the visible transverse momentum at the detector. 

We compute the one-loop QCD corrections to scalar LQ pair production at the LHC, while the LQ decays are treated at LO.
\begin{table}[tb!]\renewcommand{\arraystretch}{1.4}
	\begin{center}
		\small
			\begin{tabular}{|c|p{0.25\textwidth}|p{0.25\textwidth}|c|}
				\hline
			LQ mass &   \multicolumn{2}{c|}{$\sigma(p p \rightarrow S \bar{S})$ [fb] } & \\
				\cline{2-3}
				[GeV] & LO, $\mathcal{O}(\alpha_S^2)$   & NLO, $\mathcal{O}(\alpha_S^3)$ & K-factor\\
				\hline\hline
				1250   & $0.93^{+36.6\%}_{-25.0\%} (\pm 27.9\%)  $ & $1.448^{+11.9\%}_{-13.7\%}~(\pm 8.6\%) $ & 1.55  \\
				\hline\hline
				1300   & $0.68^{+36.7\%}_{-25.0\%} (\pm 28.6\%)  $ & $1.06^{+11.7\%}_{-13.6\%}~(\pm 9.2\%) $ & 1.56 \\
				\hline\hline
				1400   & $0.37^{+36.9\%}_{-25.1\%} (\pm 29.6\%)  $ & $0.575^{+11.7\%}_{-13.7\%}~(\pm 10.0\%)$ & 1.55\\
				\hline\hline
			\end{tabular} 
		\caption{The total LO and NLO cross sections for LQ pair production at the 14 TeV LHC, prior to their decay, are presented. The superscripts and subscripts indicate the renormalization and factorization scale uncertainties (in percentages) associated with the total cross section. The PDF uncertainty is given within the bracket. The K-factor is given in the last column.}
		\label{tab:crosssection}
	\end{center}
\end{table}
The LO cross section depends on the factorization scale, while the NLO cross section depends on both factorization and renormalization scales. To start with, we have calculated the central value of the NLO cross section after setting both factorization and renormalization scales to the central scale. We take the default choices of the {\sc MadGraph5\_aMC@NLO}\,\cite{Alwall:2014hca} for the central scale, which is half the sum of the transverse masses of all final-state particles. To estimate scale uncertainties, we vary the factorization and renormalization scales independently from half to twice the central value $\{1/2, 1, 2 \}$, generating nine distinct data sets. These cross section values for LO and NLO have been presented in Table~\ref{tab:crosssection} for various LQ masses. The superscripts and subscripts in the tables represent the uncertainty envelopes derived from these nine variations. In the NLO, the scale uncertainties are significantly reduced compared to the LO. We use {\sc NNPDF30\_LO} and {\sc NNPDF30\_NLO} PDF sets\,\cite{NNPDF:2014otw} for event generation at the LO and NLO levels, respectively. The PDF uncertainty in the total cross section is given in brackets, which is smaller than the factorization and renormalization scale uncertainty. The K-factor in the last column represents the ratio of the NLO cross section to the LO cross section corresponding to the central scale.

\subsection{Backgrounds}
The SM backgrounds that mimic the same final state are listed below.
\begin{description}[itemsep=0pt]
\item[\underline{$t\bar{t}+$jets}:]
The production of top and anti-top quark pairs at the proton-proton collider can be categorized into three scenarios based on the decay modes of the $W$-bosons produced from the decay of the top quarks. Our analysis focuses on the final state characterized by two top-like jets and significant MET without additional charged leptons.
\begin{enumerate}[itemsep=0pt]
	\item \ul{Hadronic decay of both $W$-bosons}:
	If both $W$-bosons decay hadronically, the large MET requirement and cuts on other variables that utilize MET significantly suppress this background by an order of 100 compared to the semi-leptonic decay.
	\item \ul{Leptonic decay of both $W$-bosons}:  The contribution to our desired final state is negligible if both $W$-bosons decay leptonically.
	\item \ul{Semi-leptonic decay}: The dominant contribution arises when one $W$-boson decays leptonically ($W\to \ell^+ \nu_{\ell}/\ell^- \bar{\nu}_{\ell},~\ell=e,\mu,\tau$) and the other decays hadronically. In this case, the hadronically decayed top quark produces a top-like large-$R$ jet. The second large-$R$ jet originates either from QCD radiation or from the $b$-quark of the other top quark's decay. The neutrino and the misidentified charged lepton from the leptonically decaying top quark contribute to the large MET.
\end{enumerate}
The $t\bar{t}$ background is generated with two additional QCD jets using {\sc MLM} matching\,\cite{Mangano:2006rw, Hoeche:2005vzu}. The following generation-level cuts are applied for the $t\bar{t} +$ jets and all other SM backgrounds. The minimum transverse momentum of the charged lepton is set to $p_T(\ell)=10$ GeV, while the pseudorapidity of the charged lepton is restricted to $|\eta(\ell)|\leq 3.0$. Additionally, a minimum missing transverse momentum, ${\rm MET}= 100$ GeV, and a minimum $H_T= 600$ GeV are applied. Here, $H_T$ represents the scalar sum of the transverse momentum of all visible {\sc AK04} jets in the event. These generation-level cuts are implemented because the signal is characterized by large MET and $H_T$ values (Figures \ref{AK12:met} and \ref{AK12:HT}). In the final event selection, stricter cuts on MET and $H_T$ will be applied. All the backgrounds are normalized with the available NLO, NNLO, or $N^3$LO QCD-corrected cross section\,\cite{Muselli:2015kba, Kidonakis:2015nna, Catani:2009sm, Balossini:2009sa, Campbell:2011bn}.
 
\item[\underline{Single top quark production}:]
A significant contribution to our desired final state arises from single top quark production associated with a $W$ boson ($pp\to t\, W^-$ and $pp\to \bar{t}\, W^+$). This background is generated with two additional QCD jets using {\sc MLM} matching. The primary contribution arises when one $W$ boson decays leptonically ($W\to \ell \nu_{\ell},~l=e,\mu,\tau$) while the other $W$-boson decays hadronically. The mistagged charged lepton, along with the neutrino from the $W$-boson decay, serves as the source of MET. The hadronically decaying $W$ boson or hadronically decaying top quark forms one large-$R$ jet, while the second large-$R$ jet originates from QCD radiation.

\item[\underline{Single $Z$-boson production}:]
$Z$ boson production at the LHC via quark-antiquark annihilation, followed by its subsequent decay into the invisible channel ($Z\to \nu \bar{\nu}$), contributes significantly to our final state. The neutrino pair from the $Z$ boson decay leads to large MET, while QCD radiations from the initial-state quarks mimic the large-$R$ jets. This background is generated with up to four additional QCD jets using the {\sc MLM} matching scheme.

\item[\underline{Single $W$-boson production}:]
The leptonically decaying $W^\pm$ boson is another significant background in our final state when the charged lepton escapes detection. The neutrino and the misidentified charged lepton contribute to the large MET, while QCD radiation from the initial state mimics the large-$R$ jets. This background is generated with up to four additional QCD radiations using the {\sc MLM} matching scheme.

\item[\underline{Di-boson production}:]
Di-boson production at the proton-proton collider $pp\to VV,~~(V=W^\pm, Z)$ provides a small contribution. This production can be categorized into three channels: $W^+W^-$, $W^\pm Z$, and $ZZ$. In the $W^+W^-$ channel, one $W$ boson decays leptonically, contributing to significant MET, while the other decays hadronically, forming a large-$R$ jet. For the $W^\pm Z$ channel, the $W$ boson may decay leptonically with the $Z$ boson decaying hadronically, or the $W$ boson may decay hadronically while the $Z$ boson decays invisibly. In the former case, the mistagged charged lepton and neutrino from the leptonically decaying $W$ boson contribute to MET, while the hadronically decaying $Z$ boson forms a large-$R$ jet. In the latter case, the two neutrinos from the invisible $Z$ boson decay contribute to MET. For $ZZ$ production, one $Z$ boson decays invisibly, contributing to MET, while the other decays hadronically, forming a large-$R$ jet. All these backgrounds are simulated with up to two additional QCD jets using the {\sc MLM} matching scheme, with the second large-$R$ jet originating from QCD radiation.

\end{description}
\subsection{Event Simulations}
\label{subsec:simulation}
All signal and background events are generated using {\sc MadGraph5\_aMC@NLO}\,\cite{Alwall:2014hca}. The parton-level events are then processed through {\sc Pythia8}\,\cite{Sjostrand:2001yu, Sjostrand:2014zea} for parton showering, fragmentation, and hadronization. The resulting showered events are subsequently passed through {\sc Delphes} to incorporate detector effects. Large-$R$ jets are constructed using fixed and dynamic radii to enable a comparison.
\begin{enumerate}[itemsep=0pt,topsep=1pt]
	\item \ul{Fixed radius large-$R$ jets}: Large-$R$ jets of fixed radius $R =0.8,~ 1.0,~ 1.2$, and 1.5 are constructed separately utilizing the $\mbox{anti-k}_T$\,\cite{Cacciari:2008gp} algorithm and using the particle flow candidates as implemented in {\sc Delphes} as input. The corresponding names for the jets are {\sc AK08, AK10, AK12}, and {\sc AK15}, respectively.
    \item \ul{Dynamic radius large-$R$ jets}: The dynamic radius large-$R$ jets with minimum values of radius $R_0 = 0.5$, 0.6, 0.8, and 1.0 are constructed separately utilizing the dynamic radius jet algorithm, as detailed in Subsection \ref{DR_algo}, with the particle flow candidates as implemented in {\sc Delphes} as input. The corresponding names for the jets are {\sc DRAK05, DRAK06, DRAK08}, and {\sc DRAK10}, respectively.
\end{enumerate}
Each signal and background event should have at least two large-$R$ jets. We select the constituents of the two leading large-$R$ jets and apply the soft drop algorithm to find the soft drop masses of the large-$R$ jets. Furthermore, we compute the $N$-subjettiness ratio, as detailed in Subsection \ref{N-subjettiness}. We exclusively reconstruct three subjets from the constituents of the leading large-$R$ jet and identify one as a $b$-subjet. Our custom $b$-tagging procedure is described in detail in Subsection \ref{IP_algo}. Finally, a multivariate analysis (MVA) is performed using the {\sc Delphes} generated {\sc ROOT} events, employing the adaptive BDT algorithm within the {\sc XGBoost}\,\cite{Chen:2016btl} framework.

\subsection{Event Selection}
\label{subsec:Event Selection}

The following event selection criteria are used for both signal and background processes.
\begin{enumerate}
    \item Signal and background events are selected if they contain at least two large-$R$ jets \footnote{We perform the signal and background analysis eight times, using both fixed and dynamic radius jet algorithms to form the large-$R$ jets. However, we will present our result, giving us a better statistical significance of the signal over the background. The complete set of our analysis for different algorithms and radii is $\{\rm {\sc AK08, ~AK10, ~AK12,~ AK15, ~DRAK05, ~DRAK06,~ DRAK08, ~DRAK10}  \}$.}. 
	\item Each large-$R$ jet formed using a fixed or dynamic radius algorithm should have a transverse momentum of $p_T (J_1)>200~\rm GeV$, $p_T (J_2)>200~\rm GeV$. These jets are selected within the pseudorapidity $|\eta(J_{1,2})|\leq 2.5$. $J_1$ and $J_2$ represent the leading and subleading large-$R$ jets, respectively. 
	\item A lepton veto is applied since the signal does not contain any charged leptons in the final state. Events containing an isolated lepton with $p_T (\ell)\geq 10~\rm GeV$ and $|\eta(\ell)|\leq 2.4$ ($\ell=e^\pm,\mu^\pm$) are rejected.
	\item In the signal, both large-$R$ jets originate from top quark decays. Therefore, we take the constituents of the leading large-$R$ jet and recluster them into three subjets, tagging one as a $b$-subjet. Subjet $b$-tagging within the leading large-$R$ jet ($J_1$) significantly reduces background events not associated with top quark decays.
	\item The signal exhibits a large MET due to the presence of two hard neutrinos. Therefore, we apply a large MET cut for event selection, retaining events with ${\rm MET} > 160$~GeV.
	\item To minimize the effects of jet mismeasurement, we apply azimuthal separation cuts between the large-$R$ jet and MET. Events with $\Delta\Phi (J_1, {\rm MET}) >1.0$ and $\Delta\Phi (J_2, {\rm MET})>0.2$ are selected. 
	\item A cut on the soft-drop mass of the leading ($M(J_1)$) and subleading ($M(J_2)$) large-$R$ jets is applied to reduce background events where the large-$R$ jets originate from QCD radiation or weak bosons. Events with $M(J_1)$ and $M(J_2)>110$~GeV are selected.
	\item We pair-produce two TeV-scale scalar LQs, resulting in a signal with a large $H_T$, the scalar sum of the transverse momenta of all visible {\sc AK04} jets in the event. Therefore, events are selected if $H_T>700$ GeV. The distribution of $H_T$ is shown in Figures \ref{AK12:HT} and \ref{DRAK08:HT}.
\end{enumerate}

\subsection{Signal and Background Distributions}
\label{subsec:distri}

\begin{figure}[htbp!]
	\centering
	\subfloat[] {\label{AK12:met} \includegraphics[width=0.31\textwidth]{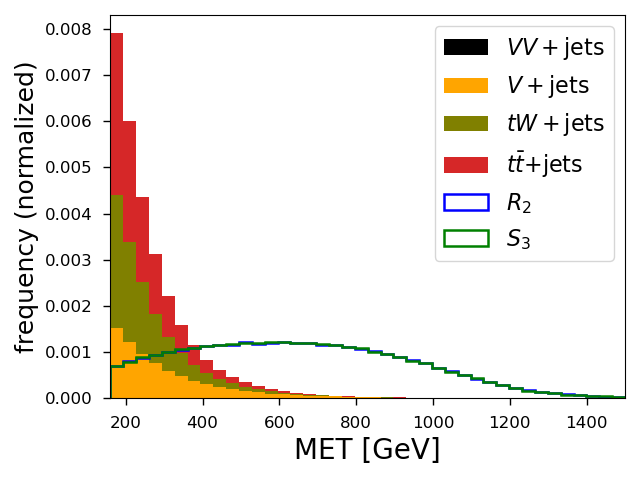}}
	\subfloat[] {\label{AK12:PTJ1} \includegraphics[width=0.31\textwidth]{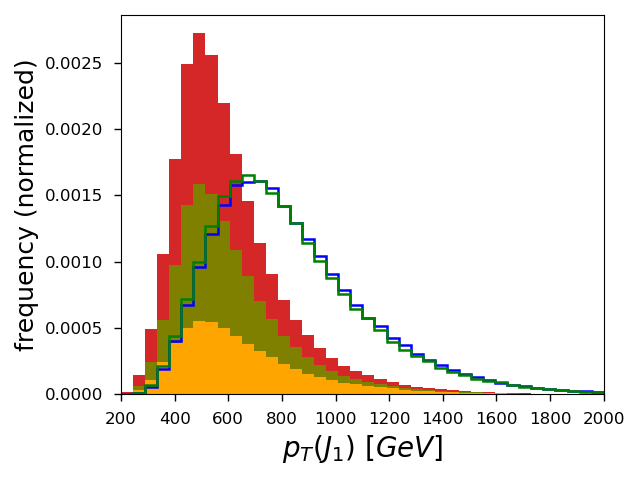}}
	\subfloat[] {\label{AK12:PTJ2} \includegraphics[width=0.31\textwidth]{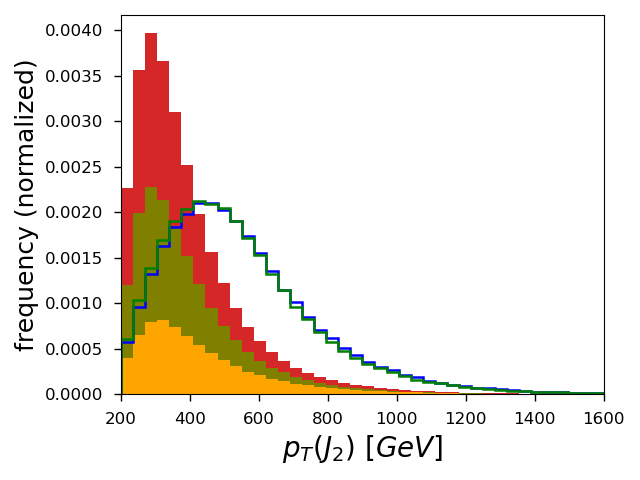}}\\
	\subfloat[] {\label{AK12:MJ1} \includegraphics[width=0.31\textwidth]{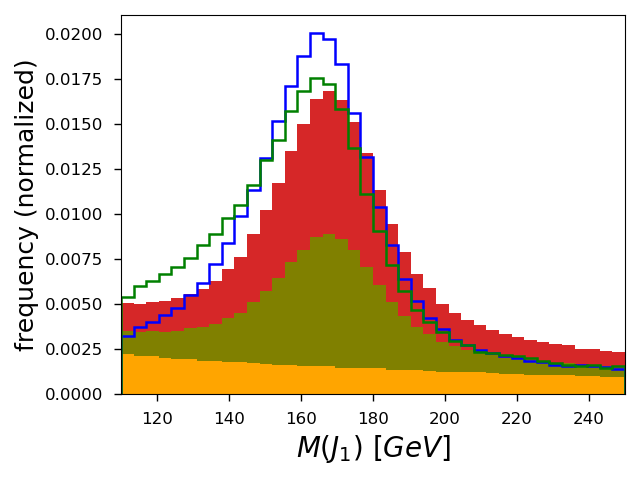}} 
	\subfloat[] {\label{AK12:MJ2} \includegraphics[width=0.31\textwidth]{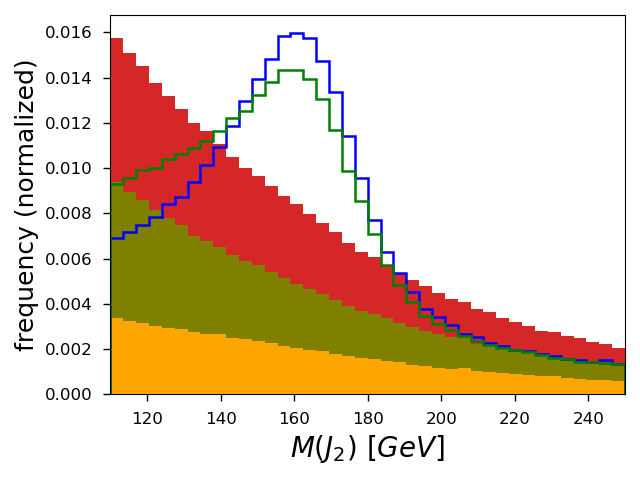}}
	\subfloat[] {\label{AK12:HT} \includegraphics[width=0.31\textwidth]{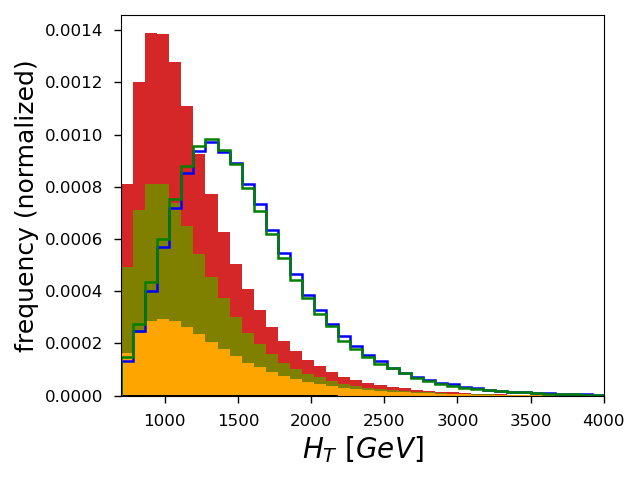}}\\
	\subfloat[] {\label{AK12:DPhi_J1ET} \includegraphics[width=0.31\textwidth]{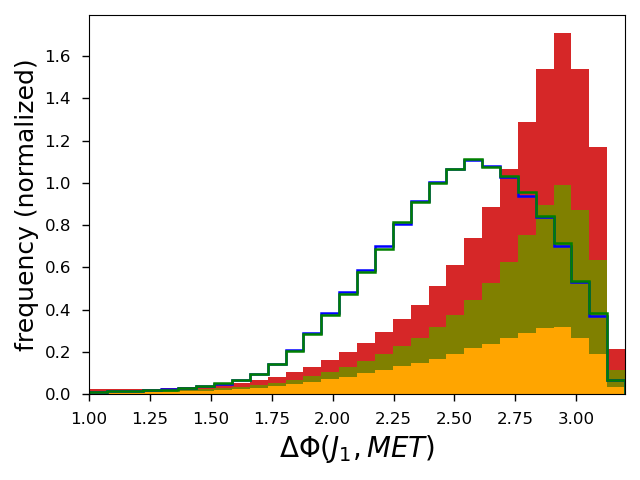}}
	\subfloat[] {\label{AK12:DPhi_J2ET} \includegraphics[width=0.31\textwidth]{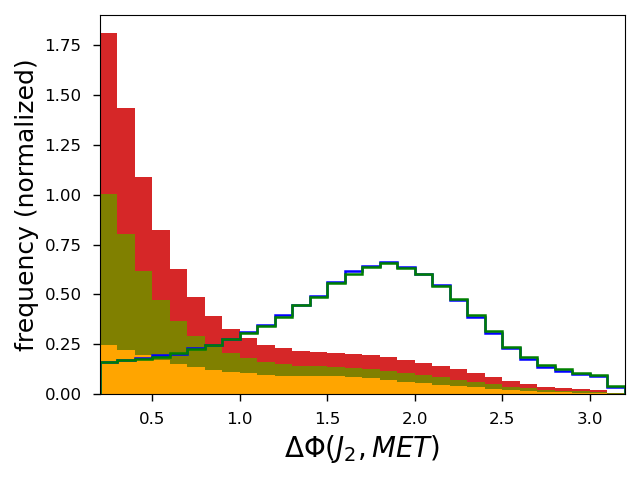}}
	\subfloat[] {\label{AK12:DRJ1J2} \includegraphics[width=0.31\textwidth]{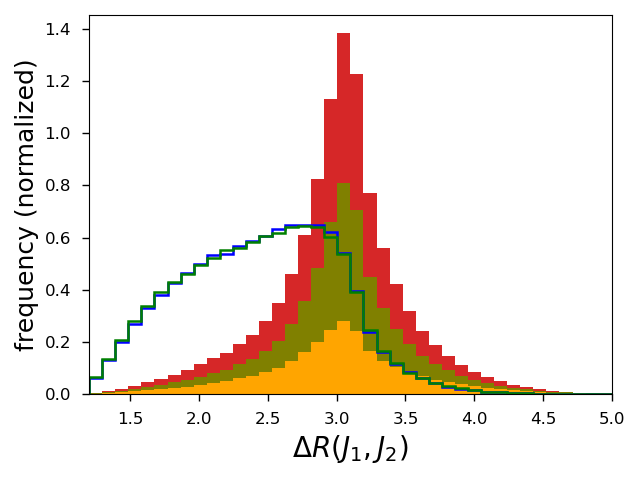}}\\
	\subfloat[] {\label{AK12:tau32_J1} \includegraphics[width=0.31\textwidth]{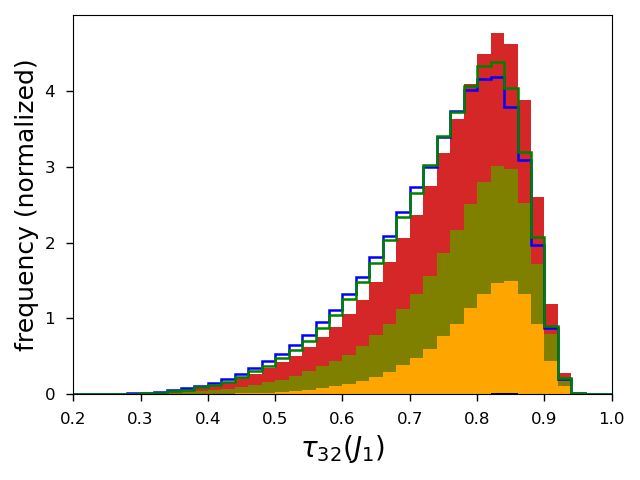}}
		\subfloat[] {\label{AK12:tau32_J2} \includegraphics[width=0.31\textwidth]{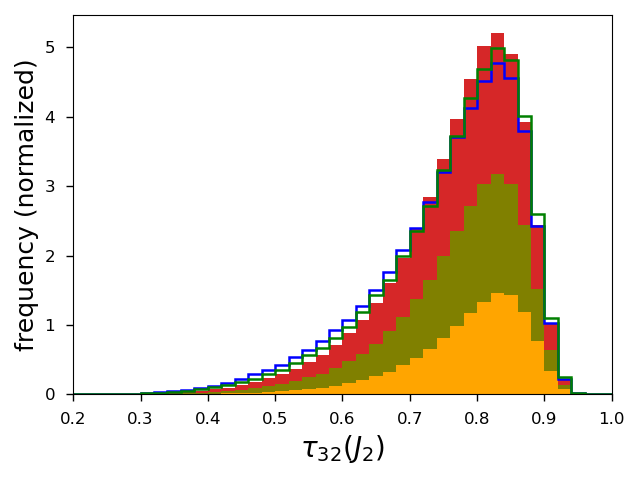}}  
	\subfloat[] {\label{AK12:Shat} \includegraphics[width=0.31\textwidth]{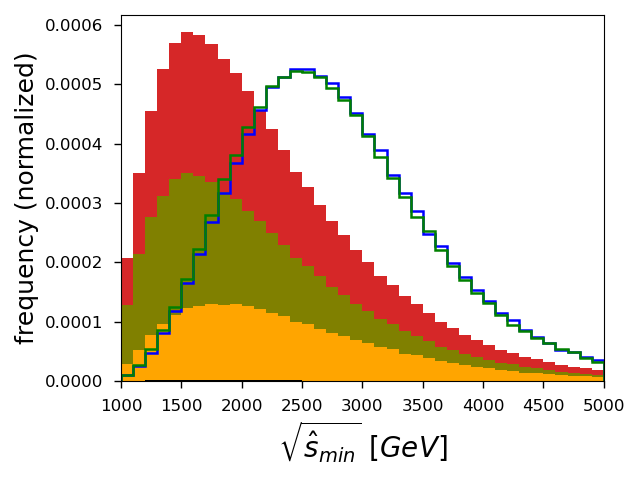}}\\
	\subfloat[] {\label{AK12:MT2} \includegraphics[width=0.31\textwidth]{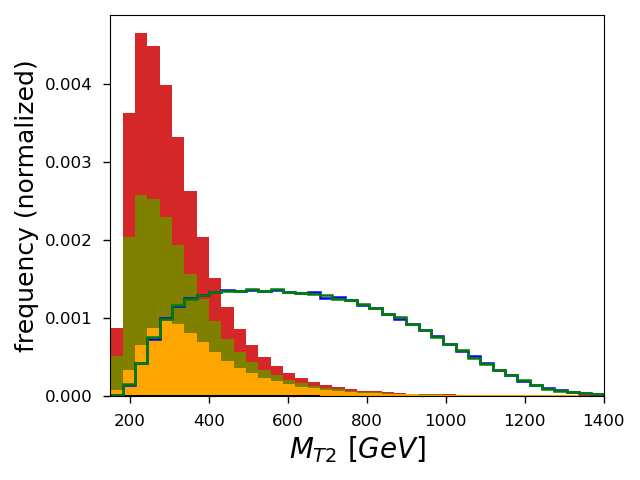}}
	\subfloat[] {\label{AK12:pTb} \includegraphics[width=0.31\textwidth]{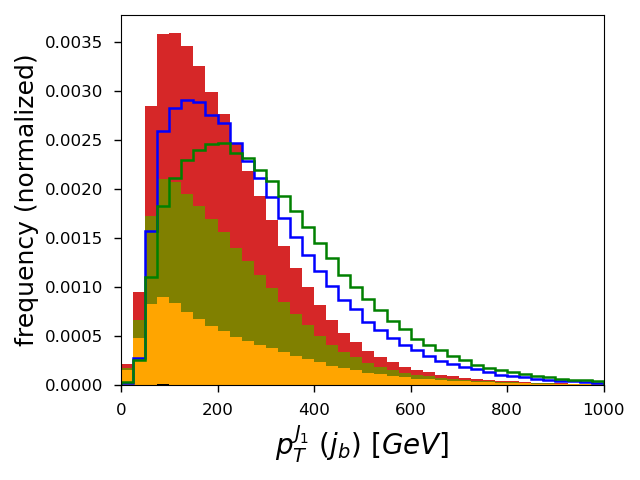}}
	\subfloat[] {\label{AK12:InvM_nonb_jets} \includegraphics[width=0.31\textwidth]{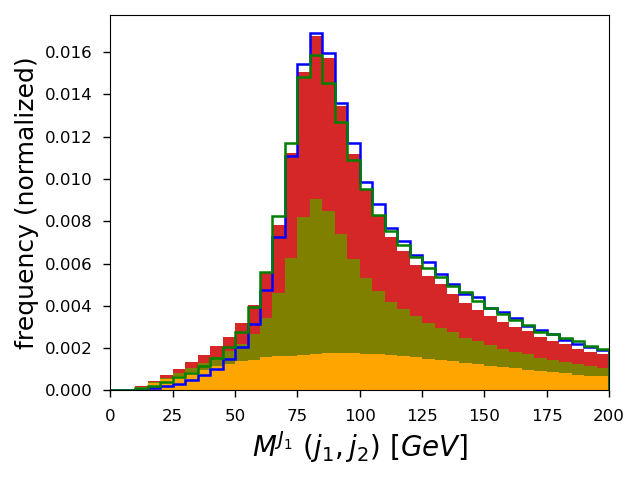}} 

	\caption{\scriptsize{Normalized bin-wise histogram for signal $S_3$ (solid green), $R_2$ (solid blue), and stacked histograms for the backgrounds after applying all the event selection cuts are presented. The distributions are shown for {\sc AK12} jets.}} 
	\label{AK12:sig_BG}
\end{figure}

After applying all the event selection criteria discussed above, the normalized signal and background distributions of various observables are presented in Figure \ref{AK12:sig_BG}. These distributions correspond to fixed-radius {\sc AK12} jets. The corresponding distributions for dynamic-radius {\sc DRAK08} jets are provided in Appendix~\ref{appen}. Most of the event distributions using fixed and dynamic radius jet algorithms
appear very similar in shape; however, their quantitative values differ, and therefore, we treat the fixed and dynamic anti-$k_t$ jet algorithms separately. The solid green and blue lines represent $S_3$ and $R_2$ signals, respectively, for an LQ mass of 1250 GeV. The contributions of individual SM background processes are represented by filled colors in the plots. The barely visible filled black region depicts the contribution of the diboson processes. Single $W$ and $Z$ channels, collectively labeled as $V+\rm jets$, are shown by the filled orange region. The filled olive and red regions illustrate contributions from the $tW+\rm jets$ and $t\bar{t}+\rm jets$ channels, respectively.

Figure \ref{AK12:met} presents the MET distributions for both signal and background. The background drops sharply at high MET, whereas the signal has a relatively uniform MET distribution within a certain range since the neutrinos from the cascade decay of TeV-scale LQs avail a larger phase space. Figures \ref{AK12:PTJ1} and \ref{AK12:PTJ2} show the transverse momentum of the leading two large-$R$ jets. The signal exhibits harder large-$R$ jets compared to the background. Figures \ref{AK12:MJ1} and \ref{AK12:MJ2} display the soft drop mass distributions of the leading two large-$R$ jets. In the $V+\rm jets$ background, no peak appears near the top quark mass, as both large-$R$ jets originate from QCD radiation. However, for $tW+\rm jets$ and $t\bar{t}+\rm jets$, a peak is observed in the leading large-$R$ jet mass distribution near the top quark mass, while the subleading large-$R$ jet lacks such a peak due to its QCD origin. In contrast, for both signal scenarios ($S_3$ and $R_2$), a distinct peak is observed for both leading and subleading large-$R$ jet masses near the top quark mass. Interestingly, the peak near the top quark mass is sharper for $R_2$ than for $S_3$, with $R_2$ having a higher event yield in this region. This difference arises from the chiral nature of the top quark in the two signals. The top quark from the decay of $S_3^{2/3}$ and $R_2^{2/3}$ LQ is left and right chiral, respectively. In the top quark’s rest frame, for $S_3$, the $b$-quark is typically aligned with the top quark’s boost direction, while the $W$ boson moves opposite to it. In contrast, for $R_2$, the $b$-quark is mainly in the opposite direction of the top quark boost, with the $W$ boson aligned with it. As a result, in the lab frame, fewer events in $S_3$ appear near the top quark mass compared to $R_2$. This is because, in $S_3$, a more substantial boost is needed to align the $W$ boson’s decay products with the $b$-quark to form a top-like large-$R$ jet. Conversely, the $b$-quark requires less boost in $R_2$ to align with the $W$ boson’s decay products and form a top-like large-$R$ jet.

The distribution of the scalar sum of the transverse momenta of all visible jets, $H_T$, is shown in Figure \ref{AK12:HT}. The signal has a significantly larger $H_T$ than the background. The azimuthal separation between the leading large-$R$ jet and MET is shown in Figure \ref{AK12:DPhi_J1ET}, while that for the subleading large-$R$ jet is in Figure \ref{AK12:DPhi_J2ET}. For the $t\bar{t}+\rm jets$ background, in the leading order, the top and anti-top quarks are in opposite directions in the $\eta-\phi$ plane. The hadronically decaying top quark produces the leading large-$R$ jet, while the leptonically decaying top quark contributes to the subleading large-$R$ jet (via the $b$-quark) and MET. Consequently, the leading large-$R$ jet and MET are nearly opposite in most events, whereas the subleading large-$R$ jet has a smaller azimuthal separation from MET. In contrast, for both signal scenarios, the cascade decay of the LQ pair produces two top quarks and two neutrinos, leading to two large-$R$ jets and MET. The neutrinos have access to a larger phase space, making the signal distribution distinctly different from the background. Consequently, only a small fraction of events exhibit a tiny azimuthal angular separation between MET and the subleading large-$R$ jet. As we will see in the next section, $\Delta \Phi (J_i, {\rm MET}),~(i=1, 2)$ are good observables for signal-background discrimination. The separation between the two large-$R$ jets in the $\eta-\phi$ plane $\Delta R(J_1,J_2)$ is illustrated in Figure \ref{AK12:DRJ1J2}.
The $N$-subjettiness ratio $\tau_{32}$ for the leading and subleading large-$R$ jets are shown in Figure \ref{AK12:tau32_J1} and \ref{AK12:tau32_J2}, respectively. It is defined as the ratio between $\tau_3$ and $\tau_2$, with their definitions provided in Subsection \ref{N-subjettiness}. Figure \ref{AK12:Shat} depicts the distribution of $\sqrt{\hat{s}_{\rm min}}$\,\cite{Konar:2008ei, Konar:2010ma}, which closely represents the minimum centre-of-mass energy of the event, with its definition given below.
\begin{eqnarray}
\sqrt{\hat{s}_{\rm min}} = \sqrt{E^2-P_z^2} + {\rm MET} \label{eq:smin}
\end{eqnarray}
Here, the four-momentum $P_\mu=(E, P_x, P_y, P_z)$ is the vector sum of all visible objects in an event. Figure \ref{AK12:Shat} indicates that signal events have higher centre-of-mass energy than the background due to the presence of two heavy LQs. The $M_{T2}$ variable\,\cite{Lester:1999tx, Konar:2009wn, Konar:2009qr, Barr:2011xt} was proposed initially to estimate the mass of a parent particle that is pair-produced at the LHC and decaying into a visible and missing particle like ours. Instead of extracting the LQ mass, we use $M_{T2}$  for signal–background discrimination. The $M_{T2}$ distribution, shown in Figure \ref{AK12:MT2}, is computed with the trial mass of the invisible particle set to zero. Large-$R$ jets originate from top quarks from heavy LQ decays in the signal, resulting in larger $M_{T2}$ values than the background. 

\begin{figure}[tb!]
	\centering
	
	\subfloat[] {\label{AK12:Eratio} \includegraphics[width=0.46\textwidth]{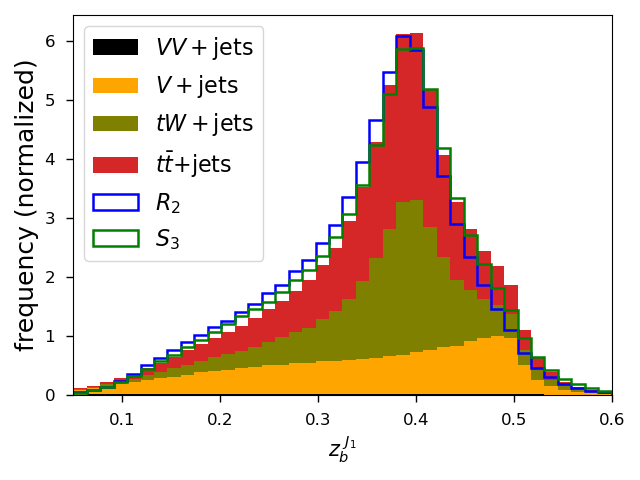}}\hspace{5mm}
	\subfloat[] {\label{AK12:cosij} \includegraphics[width=0.46\textwidth]{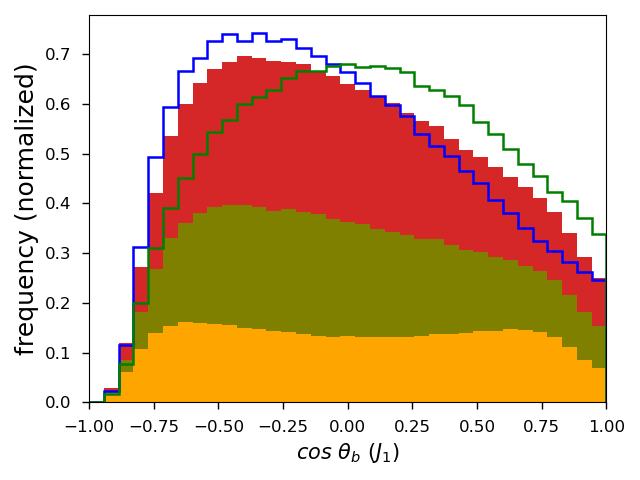}}
	
	\caption{The left panel shows the energy fraction distribution of $b$-tagged subjet in the leading large-$R$ jet, and the right panel presents the distribution of $\cos\theta_b~(J_1)$ for two signal models, $S_3$ (green) and $R_2$ (blue), along with a bin-wise stacked histograms for the backgrounds. The distributions are shown after applying all event selection cuts from Subsection \ref{subsec:Event Selection} for the leading {\sc AK12} jet.} 
	\label{Fig:Eratio&cosij}
\end{figure}

As mentioned earlier, we reclustered the constituents of the leading large-$R$ jet $J_1$ exclusively into three subjets -- one tagged as a $b$-subjet, while the other two ($j_1$ and $j_2$) are non-$b$-tagged subjets. Figure \ref{AK12:pTb} shows the transverse momentum distribution of the $b$-tagged subjet $P_T^{J_1}(j_b)$, while Figure \ref{AK12:InvM_nonb_jets} presents the invariant mass distribution of the two non-$b$-tagged subjets $M^{J_1}(j_1, j_2)$. In the top quark's rest frame, the $b$-quark in $S_3$ is predominantly aligned with the top quark's boost direction. Hence, the $b$-subjet in $S_3$ is harder than the $R_2$ model. Figure \ref{AK12:InvM_nonb_jets} shows that the invariant mass of the two non-$b$-tagged subjets peaks near the $W$-boson mass, confirming their origin from $W$ decay. This trend is also seen in the $t\bar{t}+\rm jets$ and $tW+\rm jets$ backgrounds, where the leading large-$R$ jet comes from a hadronically decaying top quark. In contrast, for the $V+\rm jets$ background, no such peak is observed near the $W$-boson mass, as the large-$R$ jets originate from QCD radiations.

Among various polarization-sensitive variables, two are shown in Figure \ref{Fig:Eratio&cosij}. Figure \ref{AK12:Eratio} presents the distribution of a variable defined as the ratio of the reconstructed $b$-subjet energy of the leading large-$R$ jet to the total reconstructed energy of the leading large-$R$ jet. As mentioned earlier, due to the different chiral nature of the top quarks arising from the decay of the LQ in different models, the $b$-subjet in the $S_3$ model is harder than in the $R_2$ model. In Figure \ref{AK12:cosij}, we plot the variable cosine of $\theta_b$, which represents the angle of the $b$-subjet of the leading large-$R$ jet relative to the boost direction of the leading large-$R$ jet in its rest frame. Since, in most events, the $b$-quark aligns with the top quark boost direction in the $S_3$ model, a more significant number of events for $S_3$ when $\cos\theta_b \geq 0$ compared to the $R_2$ model. In contrast, the $R_2$ model exhibits more events for $\cos\theta_b \leq 0$. In the next section, we will see that when distinguishing between two similar LQ signals, the $\cos\theta_b$ variable provides better discrimination power than the variable $z_b^{J_1}$. The qualitative features of these two polarization variables remain nearly identical across different fixed or dynamic jet clustering scenarios.

\begin{figure}[tb!]
	\centering
	\subfloat[] {\label{DRAK08:RJ1} \includegraphics[angle=0,width=0.45\linewidth,height=0.34\linewidth]{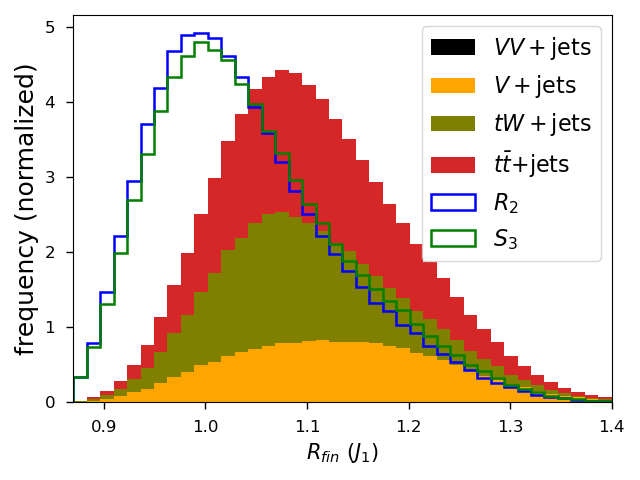}}
	\subfloat[] {\label{DRAK08:RJ2} \includegraphics[angle=0,width=0.45\linewidth,height=0.34\linewidth]{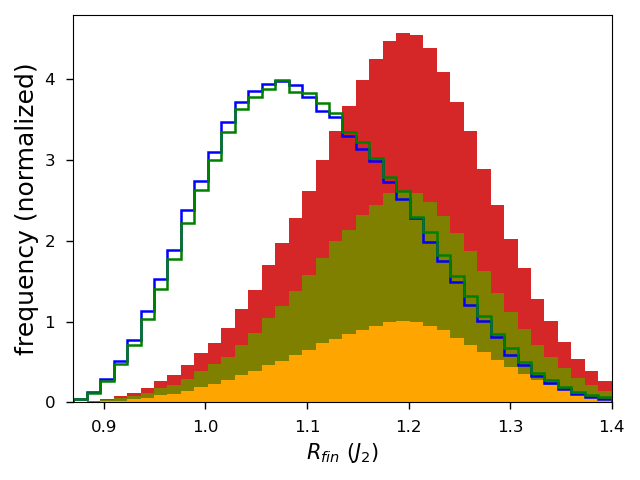}}
	\caption{Normalized distributions of the final radius of the leading and subleading large-$R$ jets for signal $S_3$ (solid green), $R_2$ (solid blue), and stacked histograms for the backgrounds after applying all the event selection cuts are presented. The distributions are shown for {\sc DRAK08} large-$R$ jets.}  
	\label{DRAK08:Rfin}
\end{figure}

Large-$R$ jets constructed using the dynamic radius jet algorithm have a radius that varies on an event-by-event basis. The normalized distributions of the leading and subleading large-$R$ jet radii are shown in Figures \ref{DRAK08:RJ1} and \ref{DRAK08:RJ2}, respectively. In Figure \ref{DRAK08:Rfin}, the large-$R$ jets are constructed using the {\sc DRAK08} algorithm, where the initial radius ($R_0$) of each large-$R$ jet is taken to be 0.8. These distributions are plotted after applying all event selection cuts, as discussed in Subsection \ref{subsec:Event Selection}, and follow the same color conventions as Figure \ref{AK12:sig_BG}. For both signal models, the large-$R$ jet radius peaks at smaller values than the background processes. This behavior is expected, as the signal large-$R$ jets originate from the decay of a TeV-scale LQ and are, therefore, more boosted than those from SM background processes. Additionally, a slight difference is observed between the two signal models due to the different chiralities of the top quarks produced in LQ decays. When using the dynamic-radius jet algorithm, these two variables exhibit good separation power for signal-background discrimination.

\subsection{Multivariate Analysis}
\label{MVA-analysis}
\begin{table}[htpb!]
	\centering
	\begin{tabular}{||c | r | p{0.68\textwidth}||} 
		\hline
		Type & Input Variable & Definition \\ [0.4ex] 
		\hline
		\multirow{4}{*}{(i)} & $M_{T2}$ & Transverse mass. \\
		 & MET & Missing transverse momentum. \\
		 & $\sqrt{\hat{s}_{\rm min}}$ & Minimum effective center of mass energy of the event [Eq.~(\ref{eq:smin})].\\
         & $H_T$ & Scalar sum of the $p_T$ of the hadrons. \\	
         \hline\hline	 		 
		\multirow{11}{*}{(ii)} & $M(J_1)$ & Soft Dropped mass of the leading large-$R$ jet ($J_1$). \\
		 & $M (J_2)$ & Soft Dropped mass of the subleading large-$R$ jet ($J_2$). \\
		 & $p_{T} ({J_1})$ & Transverse momentum of the leading large-$R$ jet. \\
		 & $p_{T} ({J_2})$ & Transverse momentum of the subleading large-$R$ jet. \\		 
		 & $E(J_1)$ & Energy of leading large-$R$ jet. \\
		 & $E(J_2)$ & Energy of subleading large-$R$ jet. \\
         & $\Delta R (J_1, J_2)$ & $\Delta R$ between leading and subleading large-$R$ jets. \\
         & $\Delta \phi (J_1, {\rm MET})$ & $\Delta \phi$ between leading large-$R$ jet and MET. \\
         & $\Delta \phi (J_2, {\rm MET})$ & $\Delta \phi$ between subleading large-$R$ jet and MET. \\
		 & $R_{\rm fin} (J_1)$ & Final acquired radius of the leading large-$R$ jet (in case of dynamic radius jet). \\
		 & $R_{\rm fin} (J_2)$ & Final acquired radius of the subleading large-$R$ jet (in case of dynamic radius jet). \\
		 \hline\hline
		 \multirow{13}{*}{(iii)}& $E^{J_1} (j_b)$ &  Energy of $b$-tagged subjet of leading large-$R$ jet. \\
		 & $E^{J_1} (j_1)$ &  Energy of leading non-$b$-subjet of leading large-$R$ jet. \\
		 & $E^{J_1} (j_2)$ &  Energy of subleading non-$b$-subjet of leading large-$R$ jet. \\
		 & $p_T^{J_1} (j_b)$ &  Transverse momentum of $b$-tagged subjet within leading large-$R$ jet.  \\
		 & $M^{J_1}~(j_1,j_2)$ &  Invariant mass of the non-$b$-subjets within leading large-$R$ jet. \\
		 & $p^{J_2}_{T} (j_1)$ & Transverse momentum of the leading subjet of subleading large-$R$ jet.\\
		 & $p^{J_2}_{T} (j_2)$ & Transverse momentum of the subleading subjet of subleading large-$R$ jet.\\
		 & $p^{J_2}_{T} (j_3)$ & Transverse momentum of the subsubleading subjet of subleading large-$R$ jet.\\
		 & $\delta E^{J_1} (j_1)$ & $|E^{J_1} (j_1) - E(J_1)/3|$ \\
		 & $\delta E^{J_1} (j_2)$ & $|E^{J_1} (j_2) - E(J_1)/3|$ \\  
		 & $\delta E^{J_1} (j_b)$ & $|E^{J_1} (j_b) - E(J_1)/3|$ \\
		 & $\tau_{32}(J_1)$ & $\tau_3/\tau_2$ of leading large-$R$ jet. \\
		 & $\tau_{32}(J_2)$ & $\tau_3/\tau_2$ of subleading large-$R$ jet. \\
		 \hline\hline
		 \multirow{2}{*}{(iv)}& $z_b^{J_1}$ & $E^{J_1}(j_b)/E(J_1)$ \\
		 & $\cos\theta_b(J_1)$ & Cosine of the angle between $b$-tagged subjet and other two combined subjet of leading large-$R$ jet in its rest frame.\\	
		\hline
	\end{tabular}
	\caption{List of feature variables used for training the BDT in the {\sc XGBoost} toolkit. The variables are categorized into four types: (i) global inclusive variables, (ii) variables using various large-$R$ jet properties, (iii) jet substructure observables, and (iv) polarization-sensitive variables using jet substructure.}
	\label{featurevar}
\end{table}

In our analysis, we explicitly recluster the constituents of each of the leading and subleading large-$R$ jets into three subjets. Additionally, one of the subjets from the leading large-$R$ jet is identified as a $b$-subjet using our custom impact parameter (IP) based algorithm (detailed in Subsection~\ref{IP_algo}). For training the BDT model, we utilize (i) several global inclusive variables, (ii) various large-$R$ jet properties, (iii) jet substructure observables, and (iv) polarization sensitive variables using jet substructure. A complete list of the feature variables and their definitions is provided in Table \ref{featurevar}. Many of these high-level variables are self-explanatory and commonly used by the HEP community. Since these polarization-sensitive variables are essential for our analysis in distinguishing different LQ models, we further discussed them in Section \ref{sec:pol_var} --- polarization-sensitive variables are not sensitive for signal-background analysis. The distributions of some of these feature variables are shown in Figure \ref{AK12:sig_BG} for a fixed radius anti-$k_t$ algorithm with $R =1.2$. The distributions for the DRAK algorithm have been presented in Appendix~\ref{appen}. The BDT classifier is trained to distinguish between two classes: the signal and the background. Since multiple SM processes can mimic the signal, we define a combined background class as a weighted sum of these contributing channels. We have two signal models, the $R_2$ and $S_3$ models, for which we optimize the BDT separately. Furthermore, we independently optimize our multivariate analysis for each clustering algorithm employed in large-$R$ jet formation. For each signal and background class, $60\%$ of the sample is randomly selected for training, while the remaining portion is used to test the BDT model.

\begin{figure}[!tb]
	\begin{center}
	\includegraphics[width=\textwidth]{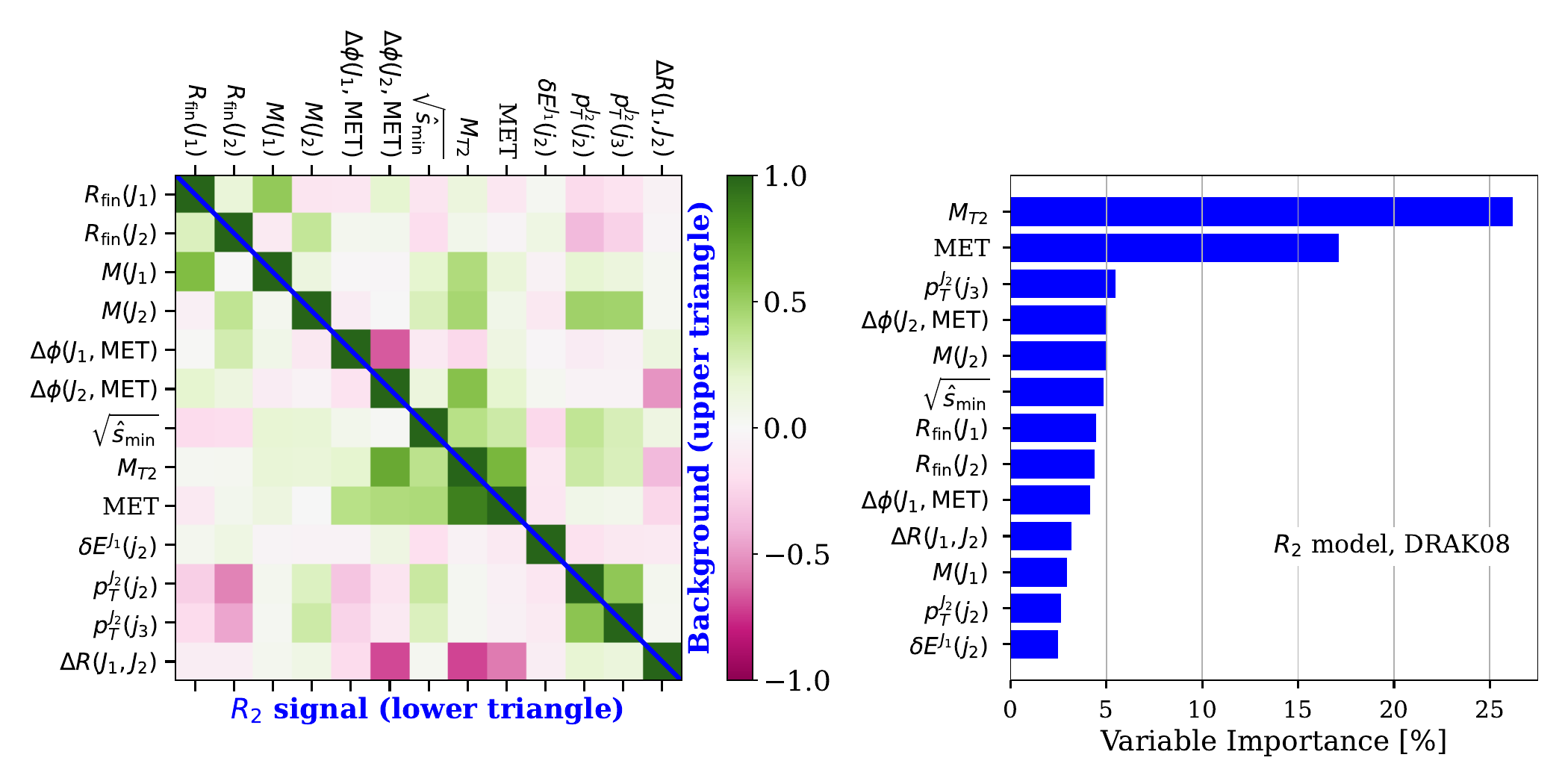}
	\end{center}
	\caption{(Left) Correlations between different variables used to train BDT for signal $R_2$ (lower triangle) and background (upper triangle). Variable ranking is shown in the right panel. The variables for large-$R$ jets and their subjets are constructed using dynamic radius anti-$k_t$ algorithm with $R_0 = 0.8$ (DRAK08).}
	\label{R2_DRAK08}
\end{figure}

\begin{figure}[!tb]
	\begin{center}
	\includegraphics[width=\textwidth]{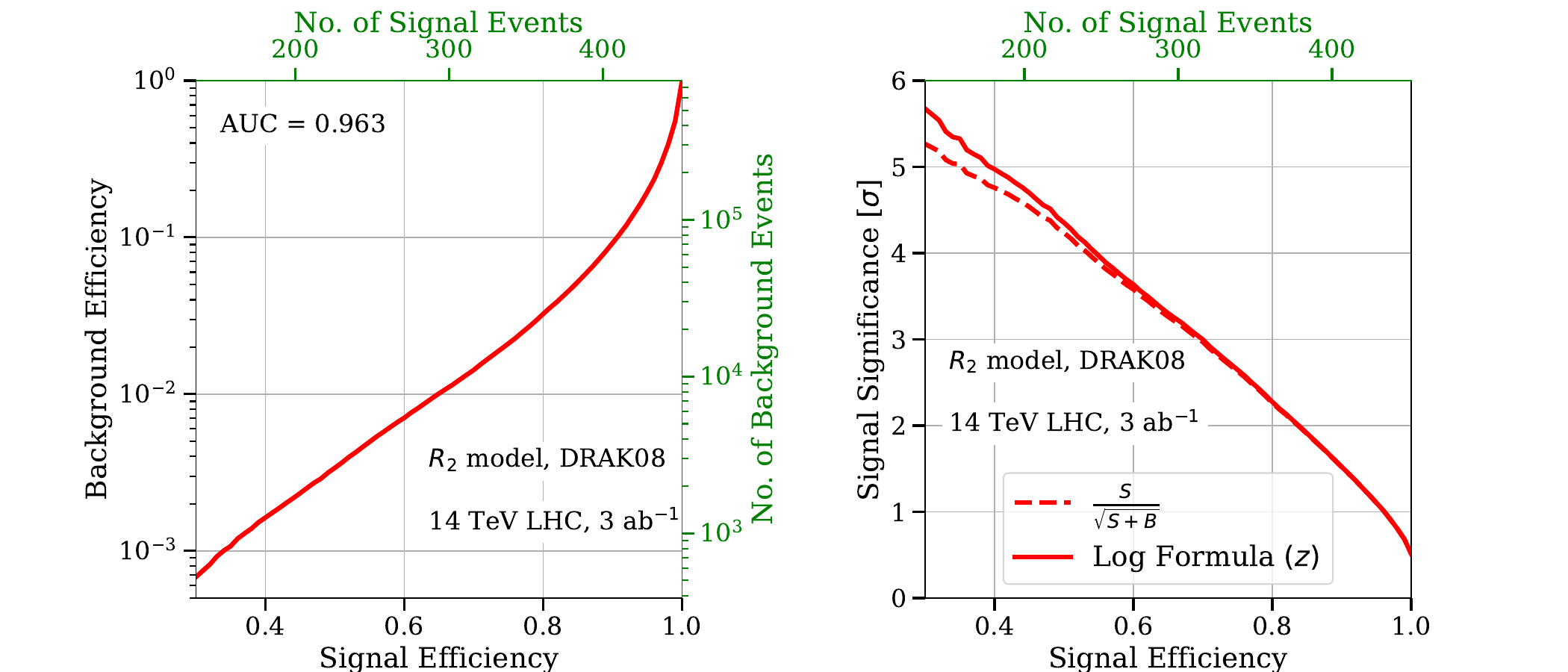}
	\end{center}
	\caption{(Left) ROC for the signal ($R_2$) and background separation. The number of signal and background events is displayed on the top and right sides of the plot for a luminosity of 3000 $\text{fb}^{-1}$. (Right) Signal significance as a function of signal efficiency. The plots are for dynamic radius anti-$k_t$ algorithm with $R_0 = 0.8$ (DRAK08).}
	\label{DRAK_R2_ROCSig}
\end{figure}

Signal and background events survived after applying all the event selection cuts detailed in Subsection~\ref{subsec:Event Selection} are used for the multivariate analysis. Some feature variables have high correlation (or anti-correlation) among themselves in both signal and background classes. Therefore, we used only those feature variables with low to moderate correlation (or anti-correlation) among themselves.
In the left panel of Figure \ref{R2_DRAK08}, the correlation matrices are shown for the $R_2$ signal-only (lower triangle) and for the background (upper triangle). The diagonal entries represent the correlation between the same variables and are equal to one. The variable importance in separating the signal $R_2$ from the background is presented in the right panel. Both plots correspond to the DRAK08 jet algorithm. Variables with an importance of less than $2\%$ are not shown in this or subsequent plots.  The Receiver Operating Characteristic (ROC) curve and its area under the curve (AUC) for $R_2$ signal versus background separation are shown in the left panel of Figure~\ref{DRAK_R2_ROCSig}. The expected number of signal and background events for an integrated luminosity of 3000 $\text{fb}^{-1}$ for a given signal and background efficiencies can be inferred from the top and right edges of the figure. The statistical significance of the signal as a function of signal efficiency is presented in the right panel. The dashed and solid lines represent the statistical significance calculated using $S/\sqrt{S+B}$ and an alternative asymptotic formula\,\cite{Cowan:2010js}
\begin{eqnarray}
	z =\sqrt{2(S+B)\log (1 + S/B)- 2S},\label{eqn:signi}
\end{eqnarray}
respectively.
In the above expressions, the symbols $S$ and $B$ represent the number of signal and background events, respectively.

\begin{figure}[!t]
	\begin{center}
		\includegraphics[width=\textwidth]{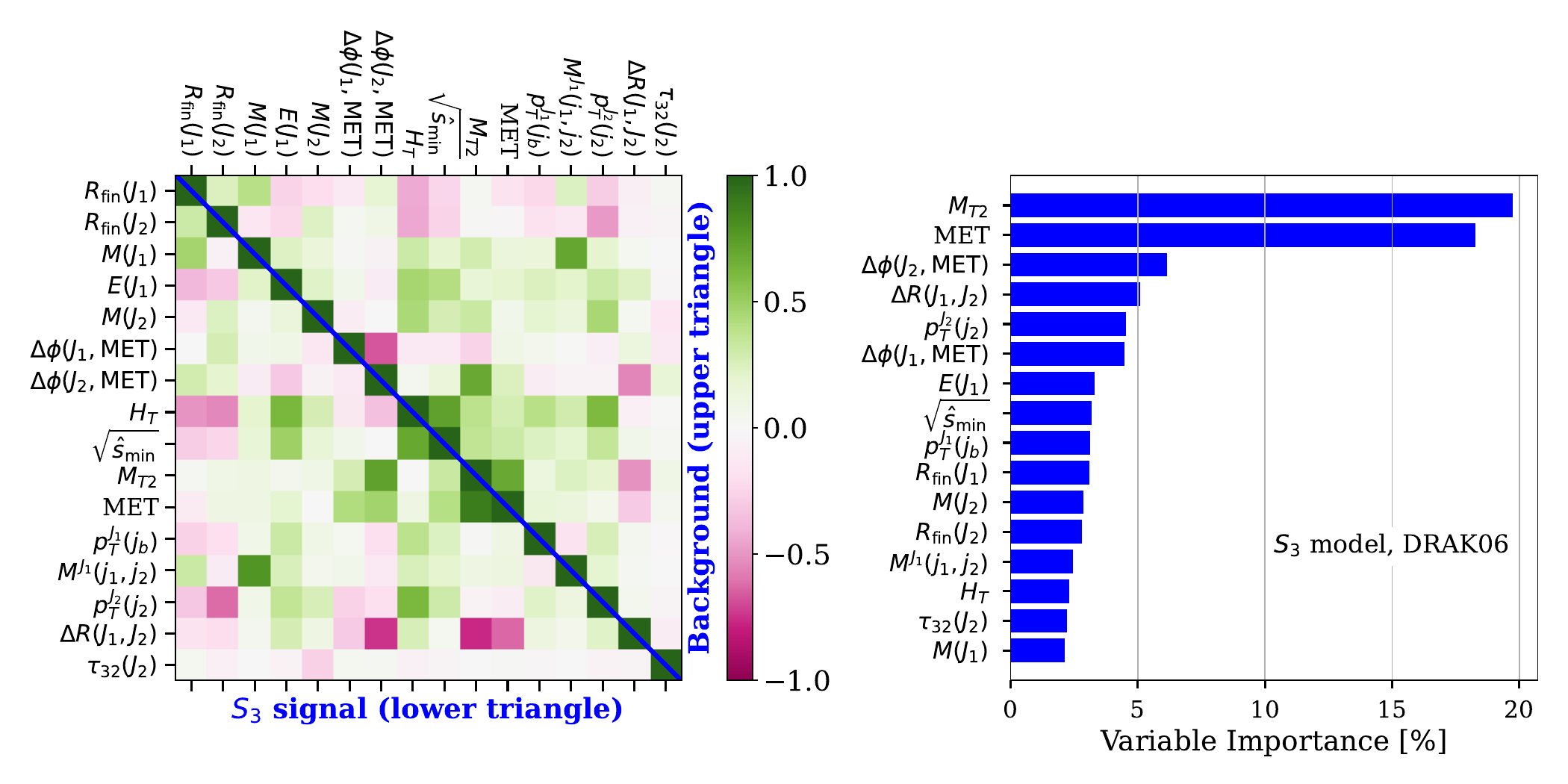}
	\end{center}
	\caption{Same as Figure~\ref{R2_DRAK08} but for $S_3$ signal model with DRAK06 jet algorithm.}
	\label{S3_DRAK06}
\end{figure}

On the other hand, to separate the signal $S_3$ from the SM background using the dynamic radius anti-$k_t$ algorithm, we train our BDT separately for different choices of $R_0$. The correlation matrix and variable ranking are shown in Figure \ref{S3_DRAK06}, where large-$R$ jets in signal $S_3$ and background are constructed using the dynamic radius anti-$k_t$ algorithm with $R_0=0.6$, referred to as DRAK06. 
\begin{figure}[!t]
	\begin{center}
		\includegraphics[width=\textwidth]{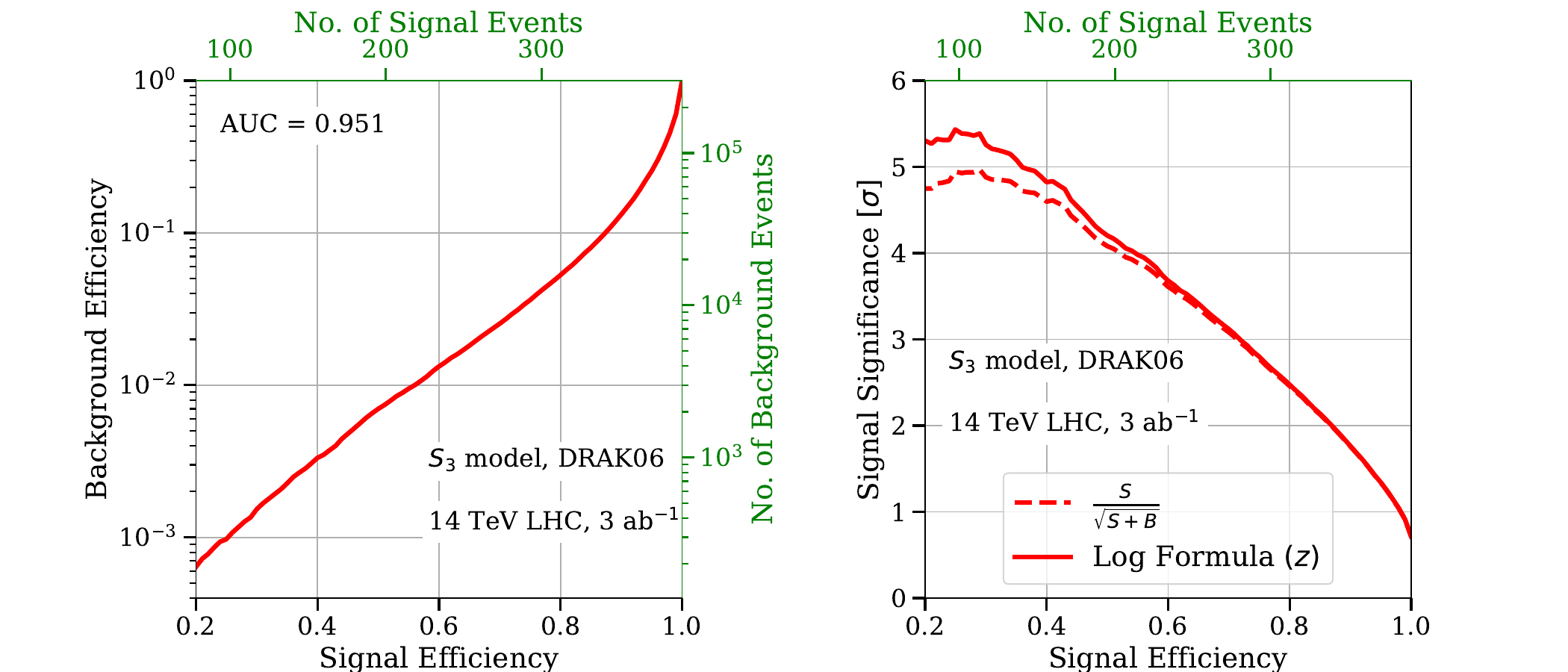}
	\end{center}
	\caption{Same as Figure \ref{DRAK_R2_ROCSig}, but the signal region is optimized for S3 model with DRAK06 jet algorithm.}
	\label{DRAK_S3_ROCSig}
\end{figure}
The ROC curve and the variation of statistical significance with signal efficiency are presented in Figure \ref{DRAK_S3_ROCSig}. 
For a signal efficiency of 0.29, we achieve a significance of $5.39\,\sigma$.

For comparison, we repeat our signal versus background analysis using the fixed-radius anti-$k_t$ algorithm. In separating the signal $R_2$ from the SM background, the correlation matrix, variable importance, ROC curve, and statistical significance are presented in the Appendix~\ref{app:BDT}, where large-$R$ jets are constructed using the fixed-radius anti-$k_t$ algorithm with $R=1.2$ (AK12).
The corresponding plots for the signal $S_3$ versus the SM background are also presented in the Appendix~\ref{app:BDT}, where large-$R$ jets are constructed using the fixed-radius anti-$k_t$ algorithm with $R=0.8$ (AK08). 

\section{Distinguishing Leptoquark Models using Polarization Variables}
\label{sec:distinguigh-models}

\subsection{Log-likelihood Ratio Test for Distinguishing $R_2$ versus $S_3$ Model}
Now that we have an acceptable signal significance, we are ready to explore the question of distinguishing the two models $R_2$ and $S_3$. This part of the query, {\it i.e.} discrimination of two prospective models, is carried out in the signal region.
In our analysis, the signal region is defined by the region containing events passing a threshold of the BDT score. We then take the statistical method of the profile likelihood estimator\,\cite{Cowan:2010js} based on a $CL_s$-type approach to try and distinguish between the two model
hypotheses\,\cite{Ghosh:2023ocz, De:2024puh}. The separation score is calculated in terms of the confidence level of excluding a hypothesis in favor of another one.

The profile likelihood estimator statistic is based on the distribution profile of a particular variable, say $u$, by which the discrimination score is calculated. Let us consider two different hypotheses $H_1$ and $H_2$, which give rise to two different distributions $\{f_1\}$ and $\{f_2\}$ corresponding to the variable $u$. Here, by $\{f_i\}$ we mean an $n$-tuple containing the bin contents of a histogrammed distribution of variable $u$. So, for this given variable, which should have a distinguishing feature or a distinct profile between two models, the likelihood ratio is calculated as
\begin{eqnarray}
	Q(H_1|H_2) = \frac{L(\{f_1\}|\{f_2\})}{L(\{f_2\}|\{f_2\})}.
\end{eqnarray}
Since, in each bin of the variable, it is essentially a number counting experiment, the likelihood function for our purposes is constructed from the Poisson distribution as the baseline distribution. Therefore, the form of the likelihood can be written as
\begin{eqnarray}
L(\{x\}|\{y\}) = \prod_{k=1}^{N} \frac{e^{-x_k} \left({x_k}\right)^{y_k}}{\Gamma\left({y_k}+1\right)}, \label{eqn:likelihood}
\end{eqnarray}
where $\Gamma$ is the gamma function. From the likelihood ratio, one then obtains the confidence level (CL), in terms of significance, at which hypothesis $H_1$ is excluded in favor of $H_2$
\begin{eqnarray}
CL_{H_1|H_2} = \sqrt{-2\,\ln Q(H_1|H_2)}
\end{eqnarray}
In the case of discrimination between model $S_3$ and $R_2$, the situation is compounded by the presence of the SM background. Therefore, the two hypotheses in consideration should be accompanied by the backgrounds, which we denote by $B+S_3$ and $B+R_2$. So, the significance of excluding model $S_3$ in favor of $R_2$ is given by 
\begin{eqnarray}
	CL_{S_3|R_2} = \sqrt{-2\,\ln Q(B+S_3|B+R_2)},
\end{eqnarray}
where both $B+S_3$ and $B+R_2$ are estimated in the signal region only. The role of $R_2$ and $S_3$ would be reversed in the case of calculating the significance of excluding model $R_2$ in favor of $S_3$. 

Now, one needs a variable by which the separation score can be achieved. Since the two models differ in their chiral structure of the couplings, polarization sensitive variables should play the role of distinguishing variables. In the following subsections, we discuss some of the important variables. 

\subsection{Polarization Variables}
\label{sec:pol_var}
The chiral structure of the coupling of the LQ should be reflected on its decay products: the top quark and the neutrino. Since the neutrinos are invisible to the detector, the study of top quark polarization (through its decay products) is crucial in determining the chiral structure of the coupling.
\subsubsection{Angular Variable}
Given the polarization of the top quark, $\mathcal{P}_t \in [-1,+1]$ the angular distribution of the decay product ($f$) of the top quark follows\,\cite{Jezabek:1988ja,Jezabek:1994zv,Godbole:2019erb}
\begin{eqnarray}
	\frac{1}{\Gamma_t} \frac{d \Gamma_t}{d \cos\theta_f} = \frac{1}{2}(1 + {\cal P}_t\,A_f \cos\theta_f),
\end{eqnarray} 
where $\theta_f$ is the angle between the top quark propagation direction and the decay product $f$ in the top quark rest frame. The polarization value $-1$ represents a completely left-handed state, $+1$ represents a completely right-handed state, while other values correspond to mixed states. For the decay product $f$, $\kappa_f$ represents its spin analyzing power. We are interested in the hadronic decays of the top quark and the decay products for top quarks are $b$, $u/c$, and $\bar d/\bar{s}$. The corresponding spin analyzing powers of these decay products are $A_b = -0.4$, $A_{u/c} = -0.3$, and $A_{\bar d/\bar s} = 1.0$. The spin analyzing power of the $W$ boson is opposite to that of the $b$ quark, {\it i.e.}, $A_W = -A_b = 0.4$.

\begin{figure}[!t]
	\centering
	\includegraphics[width=0.96\textwidth]{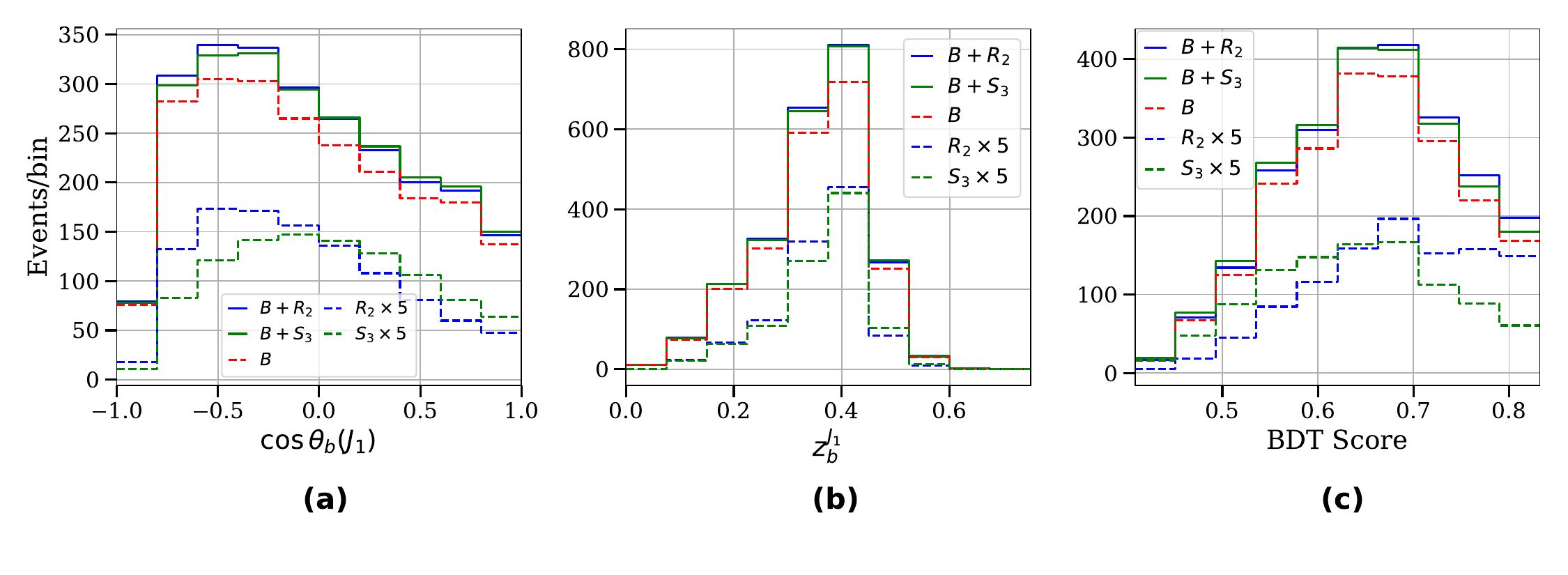}
	\caption{Distributions of polarization sensitive variables in the signal region (optimized for $R_2$ model with DRAK08 jet algorithm) for $R_2$, $S_3$, $B$, $B+R_2$, and $B+S_3$, where $B$ represents the SM backgrounds. For the signals, the histogram values are multiplied by a factor of 5 to better show the contrast.  The `BDT score' variable is an optimized variable combining more than one variable (see text). The distributions are shown for an integrated luminosity of 3000~fb$^{-1}$ at 14~TeV HL-LHC.}
	\label{fig:excluR2DRAK08}
\end{figure}

\begin{figure}[!h]
	\centering
	\includegraphics[width=0.96\textwidth]{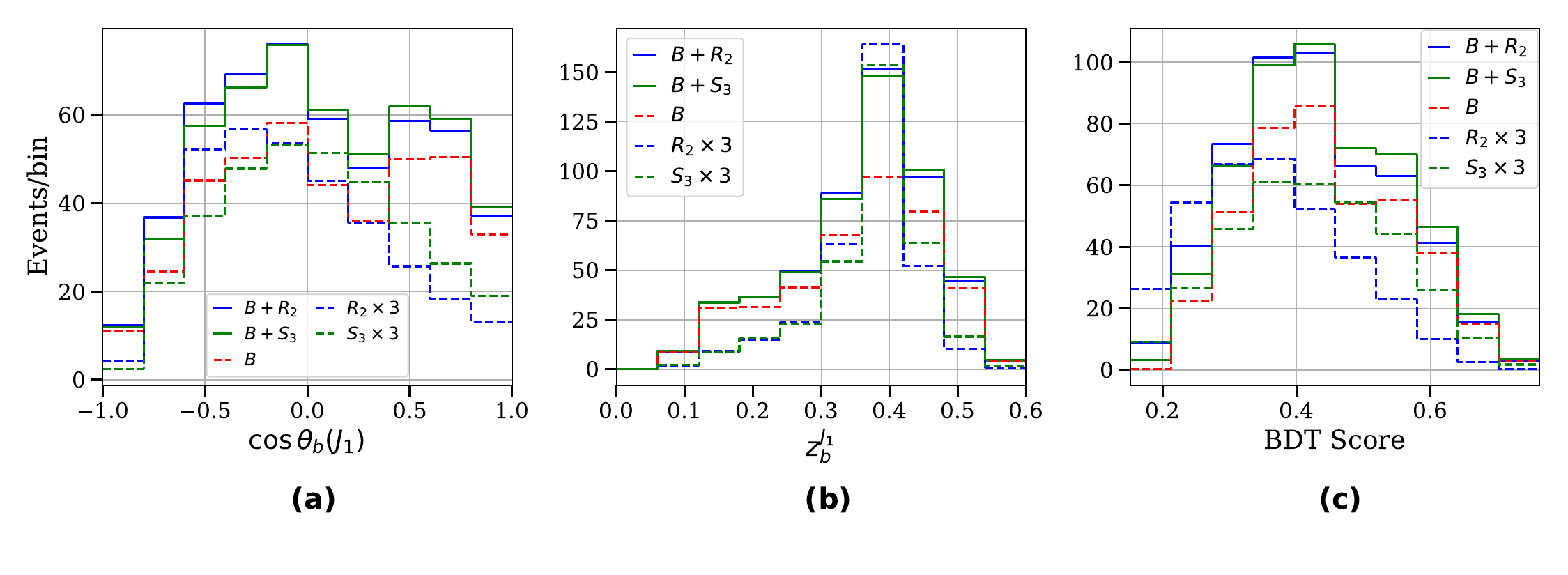}
	\caption{Same as Figure \ref{fig:excluR2DRAK08}, but the signal region is  optimized for $S_3$ model with DRAK06 jet algorithm. Here, the signal histogram has been scaled by a factor of 3 to enhance the visual contrast. }
	\label{fig:excluS3DRAK06}
\end{figure}

In the collider, we need to work with the reconstructed-level objects as these partons are seen as jets or subjets after showering and hadronization. Although the $\bar d/\bar s$ quark has the highest spin analyzing power, especially in the highly hadronic environment, tagging $\bar d$ or $\bar s$ is not possible. However, $b$-jet tagging is well developed, albeit with a small reduction in tagging efficiency. In the case of a $b$-subjet, our modification to account for this has been discussed in Section~\ref{IP_algo}. Therefore, we will be focusing on the variable $\cos\theta_b$ of the leading boosted top jet. The distributions of the variable $\cos\theta_b (J_1)$ in the signal region, at the reconstructed level, have been shown in Figure \ref{fig:excluR2DRAK08}(a) for the jet algorithm DRAK08 and in Figure \ref{fig:excluS3DRAK06}(a) for the jet algorithm DRAK06. In both figures, the dashed lines represent the distributions for the signal models $R_2$ (blue) and $S_3$ (green) while the corresponding solid lines represent the distributions with the SM background and signal combined. The dashed red line represents the SM background. To highlight the contrast between the two signal models, the dashed histograms are multiplied by a factor of 5 and 3 in Figure \ref{fig:excluR2DRAK08}(a) and in Figure \ref{fig:excluS3DRAK06}(a), respectively. We can clearly see differences in the distributions between $S_3$ and $R_2$. Since the LQ in the $R_2$ model tends to be right-handed, the distribution is shifted towards $\cos\theta_b(J_1)=-1$ compared to the $S_3$ model. However, as expected, the distribution when combined with the background is smeared. As before, the plots corresponding to the fixed radius algorithms (AK12 and AK08) have been moved to Appendix~\ref{app:polvars}.

\subsubsection{Energy Fraction}
Another variable that captures the polarization information of the top quark and can be calculated in the lab frame is the energy fraction carried by the $b$-quark defined simply by $z_b = E_b/E_t$, where $E_b$ and $E_t$ are the energies of the $b$ and $t$ quarks, respectively\,\cite{Mandal:2015lca}. The distribution of $z_b$, in the massless $b$-quark limit, follows the distribution
\begin{eqnarray}
	\frac{1}{\Gamma_t} \frac{d \Gamma_t}{d z_b} = \frac{m_t^2}{\beta_t(m_t^2-m_W^2)}\left[1-\frac{\mathcal{P}_t \kappa_b}{\beta_t}\left(\frac{2m_t^2}{m_t^2-m_W^2}z_b - 1\right)\right]
\end{eqnarray}
for a top quark with a Lorentz boost $\beta_t$. This distribution is correlated to the variable $\cos\theta_b$. This variable ranges between $[0,(1-m_W^2/m_t^2)]$ and the distribution tends toward larger values for a right-handed top quark compared to a left-handed one. This feature can be seen in Figures \ref{fig:excluR2DRAK08}(b) and \ref{fig:excluS3DRAK06}(b), where we have shown the energy fraction variable for the leading large-$R$ jet. The representation and color conventions are the same as those used for $\cos\theta_b$. In this case also, the difference becomes less prominent when the background distribution is taken along with the signals. The plots corresponding to the fixed radius algorithms have been shown in Appendix~\ref{app:polvars}.

\subsubsection{Combined Variable: BDT Score}
A combination of variables is more useful in separating the two signal models, since it has the capacity to exploit the power of all the variables involved. We have used the multivariate technique BDT to make a combined variable. We have adopted a similar procedure for this BDT network as explained in Section~\ref{MVA-analysis}. The differences with respect to the above networks are in (1) the choice of variables, and (2) the dataset used for classification. In this part of the network, we have used only the second set of variables (separated by two horizontal lines) listed in Table \ref{featurevar}. For the dataset, the events of the two different models have been used, {i.e.}, without SM background. After the training with optimized hyperparameters, the BDT classification score has been used as the combined variable. Four separate BDT networks have been trained for DRAK08, DRAK06, AK12, and AK08. The distribution corresponding to these two combined variables are shown in Figures \ref{fig:excluR2DRAK08}(c) and \ref{fig:excluS3DRAK06}(c) for DRAK08 and for DRAK06, respectively, while the plots corresponding to the fixed radius algorithms have been moved to Appendix~\ref{app:polvars}. One can clearly see better separability between $R_2$ and $S_3$ models in these combined variables. The color conventions for these plots are the same as the other two variables. We now proceed to the next section to see the separation power of the two models with respect to the three variables discussed.

\section{Results and Discussions}
\label{sec:results}
We now present our findings in terms of (a) separating each of two signals from the SM background in terms of signal significance as defined in Eq.~\ref{eqn:signi}, and (b) separating $R_2$ from $S_3$ and {\it vice versa}, following the procedure described in Section~\ref{sec:distinguigh-models}. We first have optimized the radius parameter of the jet clustering algorithms and the BDT network for signal-background separation to achieve the best signal significance for the $R_2$ and $S_3$ models separately. For each case, we select a working point of the BDT score threshold value for which the highest signal significance is achieved. For the $R_2$ model, we find that the optimal radius parameter is $R = 1.2$ for the fixed radius anti-$k_t$ and $R_0 = 0.8$ for the dynamic radius anti-$k_t$ algorithm. For both jet algorithms, the signal efficiency ($\epsilon_S$) and background acceptance ($\epsilon_B$) corresponding to the best signal significance, along with the number of signal events and number of background events for an integrated luminosity of 3000 $\rm fb^{-1}$, are tabulated in Table \ref{tab:signi_r2} for $R_2$ as the signal model. The corresponding signal significances that can be achieved are $4.83\,\sigma$ and $4.52\,\sigma$ for AK12 and DRAK08 jet algorithms. We present in the last three columns of Table \ref{tab:signi_r2}, the confidence level of excluding the $S_3$ model in favor of $R_2$ corresponding to the variables $z_b^{J_1}$, $\cos\theta_b(J_1)$, and `BDT score' as a combined variable. It is evident that among these, the separation power of the energy fraction variable $z_b^{J_1}$ is the lowest, that of the angular variable is moderately better, and the corresponding confidence levels are $0.5\,\sigma$ and $1.1\,\sigma$, respectively. However, the BDT score can achieve an exclusion significance of $2.29\,\sigma$ for AK12 and $2.12\,\sigma$ for DRAK08.

\begin{table}[!t]
\begin{tabular}{|c|c|c|c|c|c|c|}
\hline
\multirow{2}{40pt}{\centering Large-R Jet} & \multirow{2}{*}{S ($\epsilon_S$)}     & \multirow{2}{*}{B ($\epsilon_B$)}      & \multirow{2}{55pt}{Signal Significance}    & \multicolumn{3}{p{120pt}|}{\centering $CL_{S_3|R_2}$} \\[1.0pt]
\cline{5-7}    &                    &                      &  & $z_b^{J_1}$ & $\cos\theta_b (J_1)$ & BDT score       \\ \hline
AK12        &  206 (0.45)       &  1753 ($1.76\times 10^{-3}$)                &  4.83   & 0.51  & 1.18 & 2.29                     \\ \hline
DRAK08        &   217 (0.48)     &   2231 ($2.89\times 10^{-3}$)             & 4.52    & 0.52 & 1.08 & 2.12         \\ \hline
\end{tabular}
\caption{Optimal working points corresponding to the best signal significance for model $R_2$. The last three columns represent the significance of excluding $S_3$ in favour of $R_2$ using three different variables for an integrated luminosity of 3000 $\rm fb^{-1}$. The values have been presented for AK12 and DRAK08 jet algorithms.
}
\label{tab:signi_r2}
\end{table}
A similar analysis has been performed considering the $S_3$ model as our signal model. In this case, the optimal radius choices for fixed radius and dynamic radius algorithms are 0.8 and 0.6, respectively, for the signal significance analysis. For these two jet algorithms, the signal efficiency ($\epsilon_S$) and background acceptance ($\epsilon_B$) after putting a cut on the BDT score to achieve best signal significance are given in Table \ref{tab:signi_s3}. The number of events for signal and background at the specified efficiencies are also provided in the same table. The corresponding signal significances are $5.37\,\sigma$ and $5.39\,\sigma$ considering jet algorithms AK08 and DRAK06, respectively. As before, we then performed an analysis to obtain the significance of excluding $R_2$ over the $S_3$ model using $z_b^{J_1}$, $\cos\theta_b(J_1)$, combined BDT score variable. The variables help us achieve exclusion significances of $0.69\,\sigma$, $1.4\,\sigma$, and $3.23\,\sigma$ for the DRAK06 jet algorithm, corresponding to the variables $z_b^{J_1}$, $\cos\theta_b(J_1)$, and the combined BDT score, respectively. In the case of $S_3$, the dynamic radius algorithm performs marginally better than the fixed radius one.

\begin{table}[!t]
\begin{center}
\begin{tabular}{|c|c|c|c|c|c|c|}
\hline
\multirow{2}{40pt}{\centering Large-R Jet} & \multirow{2}{*}{S ($\epsilon_S$)}     & \multirow{2}{*}{B ($\epsilon_B$)}      & \multirow{2}{55pt}{Signal Significance}    & \multicolumn{3}{p{120pt}|}{\centering $CL_{R_2|S_3}$} \\[1.0pt]
\cline{5-7}    &                    &                      &  & $z_b^{J_1}$ & $\cos\theta_b (J_1)$ & BDT score       \\ \hline
AK08        &  149 (0.39)      &  722 ($2.67\times 10^{-3}$)  & 5.37 & 0.63  & 1.35 &  3.06         \\ \hline
DRAK06      &   113 (0.29)     &  404  ($1.35\times 10^{-3}$) & 5.39 & 0.69  & 1.40  &  3.23   \\ \hline
\end{tabular}
\end{center}
\caption{Optimal working points corresponding to the best signal significance for model $S_3$. The last three columns represent the significance of excluding $R_2$ in favour of $S_3$ using three different variables for an integrated luminosity of 3000 $\rm fb^{-1}$. The values have been presented for AK08 and DRAK06 jet algorithms.
}
\label{tab:signi_s3}
\end{table}

Overall, Tables \ref{tab:signi_r2} and \ref{tab:signi_s3} show that a comparatively larger jet radius (both for fixed and dynamic radius algorithms) gives better statistical significance of the signal $R_2$ over the SM background. In contrast, a smaller jet radius improves the significance of the $S_3$ signal. This difference arises from the kinematics of the top quark decay. In the rest frame of the top quark, in the $R_2$ model, most of the $W$ bosons are in the direction of the top quark boost, and the $b$-quark is opposite to it. The opposite happens to the $S_3$ model; refer to Figure \ref{AK12:cosij} which is computed after the reconstruction of large-$R$ jets and $b$-subjet. Consequently, the $R_2$ model produces more three-pronged large-$R$ jets in the lab frame, whereas the $S_3$ model favors two-pronged jets. Figures \ref{AK12:MJ1} and \ref{AK12:MJ2} show more $R_2$ events clustering near the top quark mass. Despite identical final states, pair production processes, and partonic cross sections, their distinct chiralities drive different jet substructures. In the $R_2$ model, most events form three-pronged large-$R$ jets, whereas in the $S_3$ model, two-pronged structures dominate. This explains why a larger jet radius improves sensitivity for $R_2$ while a smaller radius is optimal for $S_3$.

\section{Summary and Conclusion}
\label{sec:conclsn}
The models containing leptoquarks remain one of the intriguing extensions beyond the SM. In this work, we have studied the pair production of scalar leptoquarks, which couple primarily to the third-generation lepton and quark sectors. The study is performed in the decay mode in which each leptoquark subsequently decays to a top quark and a neutrino. For heavy leptoquarks beyond 1~TeV, the resultant top quarks become boosted, leading to a 3-pronged large-$R$ jet and large MET from the undetectable neutrinos. Therefore, we have particularly looked at the signal with two large radius jets and large MET in two different leptoquark models, namely the $R_2$ and $S_3$ models. We have used traditional fixed radius anti-$k_T$ and dynamic radius anti-$k_T$ jet algorithms to compare their performances. We make use of the jet substructure techniques, namely Soft Drop and $N$-subjettiness, to find the subjet inside the boosted top jets and to tag $b$-subjet within the leading large-$R$ jet, a key feature of top-like large-$R$ jets. Additionally, by adopting multivariate analysis through BDT, we could enhance the sensitivity of our study. We have observed that the optimal choices for radius parameters of the jet algorithms are different for the two leptoquark models to achieve the best signal significance; a smaller radius is preferable for the $S_3$ model, and a larger radius is preferable for the $R_2$ model. Considering a luminosity of 3000~fb$^{-1}$ at the 14 TeV HL-LHC  leptoquark mass of 1.25~TeV, we could achieve a significance of $4.83\,\sigma$ for the $R_2$ model and $5.39\,\sigma$ for the $S_3$ model.

We then investigated the distinguishability of the two models, which differ in their chiral structure of the coupling to the top quark and the neutrino. This has been performed using polarization sensitive variables such as the angular variable $\cos\theta_b$ or the energy fraction of the $b$-subjet to the top jet ($E_b/E_t$). We have used a $CL_s$-type profile likelihood estimator to find confidence levels of discriminating one model from another in terms of exclusion significance. The most sensitive variable is $\cos\theta_b$, which yields exclusion significances $1.1\,\sigma$ for excluding $R_2$ over $S_3$ and $1.4\,\sigma$ for excluding $S_3$ over $R_2$ in the signal regions. We then used another BDT network, trained with multiple variables against the two signal models, to obtain a BDT classification score as a combined variable, which outperforms any single polarization-sensitive variable. With this combined variable, exclusion significances are $2.29\,\sigma$ for excluding $R_2$ in favor of $S_3$ and $3.23\,\sigma$ for excluding $S_3$ in favor of $R_2$. In conclusion, our study suggests that third-generation scalar leptoquark pair production can be probed at the HL-LHC in the two large-$R$ jets plus MET channel with a significance exceeding $5\,\sigma$. Simultaneously, the two leptoquark models can be distinguished with a significance greater than $3\,\sigma$.

\section*{Acknowledgements}
The authors acknowledge the XVII Workshop in High Energy Physics Phenomenology (WHEPP XVII) organized at IIT Gandhinagar for hospitality and the environment for engaging discussions during which the project was initiated. The authors also acknowledge the Param Vikram-1000 High-Performance Computing Cluster and the TDP project resources of the Physical Research Laboratory (PRL) for various computational works.

\section*{Data Availability Statement}
This study uses simulated data from standard HEP packages. The details of generating the simulated data supporting these findings are available within this article. 

\appendix
\section{Event Variables for Dynamic Radius Jet Algorithm}
\label{appen}
\begin{figure}[h!]
	\centering
	\subfloat[] {\label{DRAK08:met} \includegraphics[width=0.31\textwidth]{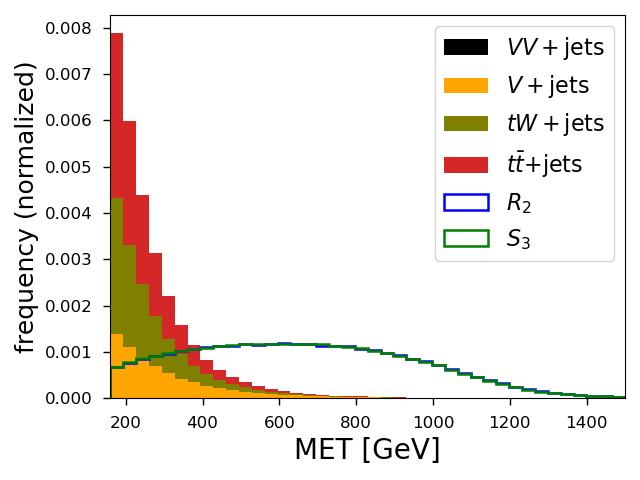}}
	\subfloat[] {\label{DRAK08:PTJ1} \includegraphics[width=0.31\textwidth]{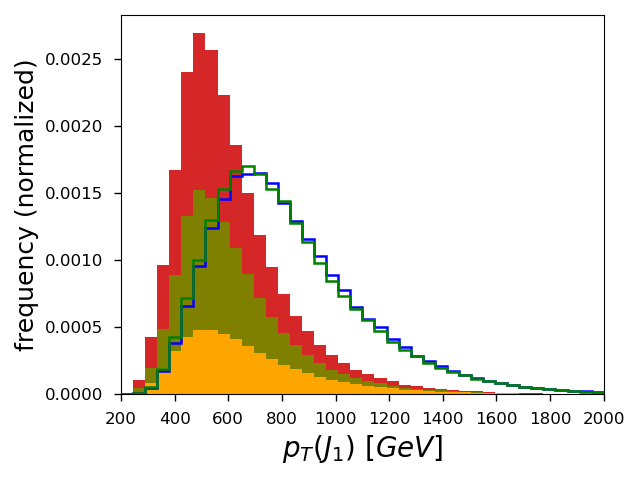}}
	\subfloat[] {\label{DRAK08:PTJ2} \includegraphics[width=0.31\textwidth]{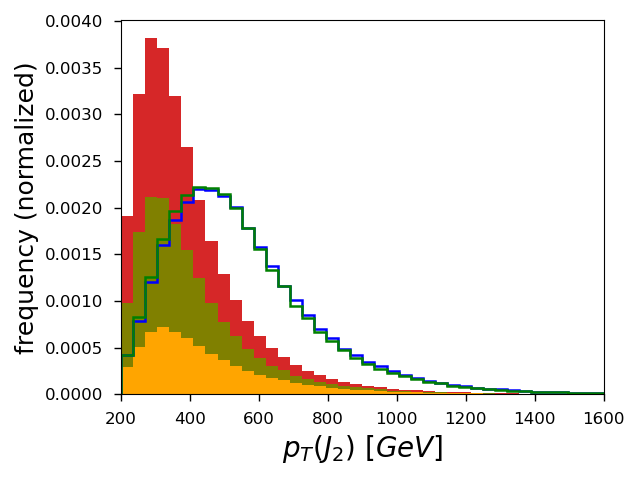}}\\
	\subfloat[] {\label{DRAK08:MJ1} \includegraphics[width=0.31\textwidth]{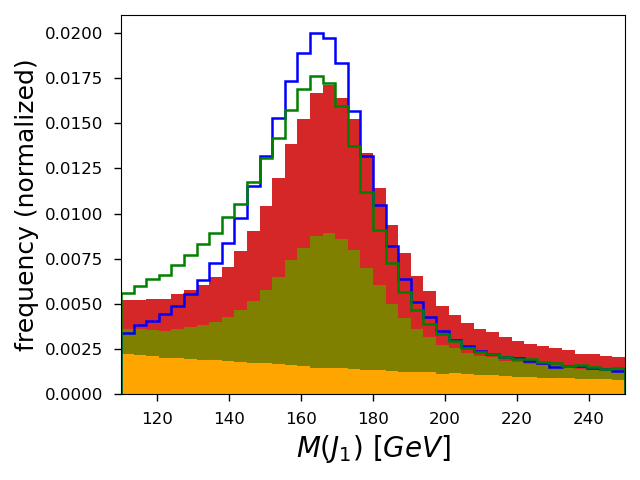}} 
	\subfloat[] {\label{DRAK08:MJ2} \includegraphics[width=0.31\textwidth]{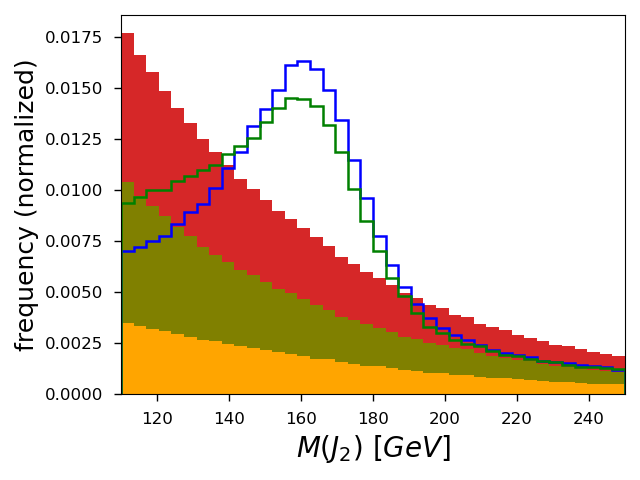}}
	\subfloat[] {\label{DRAK08:HT} \includegraphics[width=0.31\textwidth]{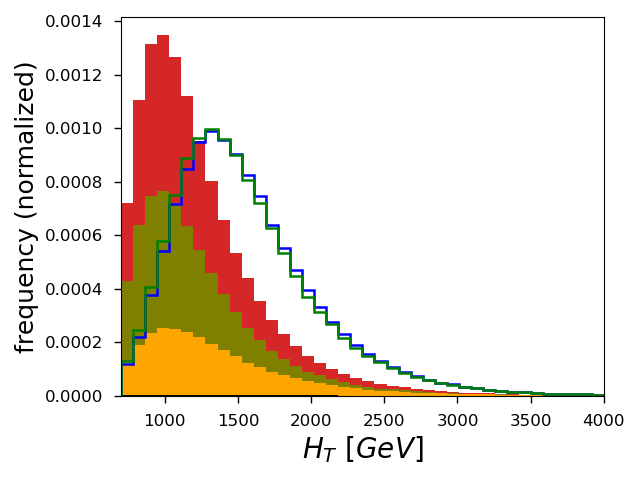}}\\
	\subfloat[] {\label{DRAK08:DPhi_J1ET} \includegraphics[width=0.31\textwidth]{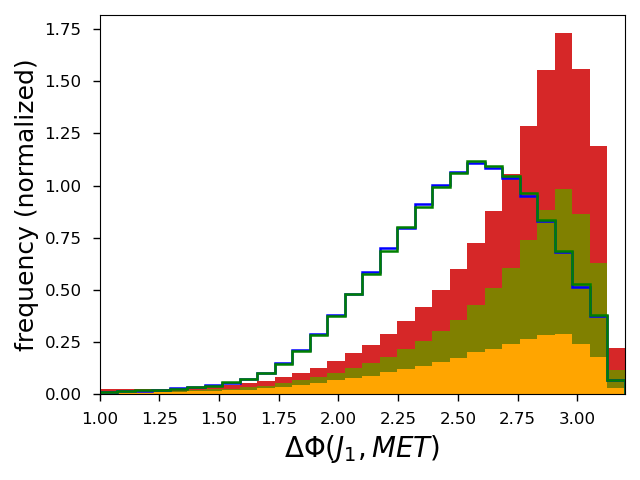}}
	\subfloat[] {\label{DRAK08:DPhi_J2ET} \includegraphics[width=0.31\textwidth]{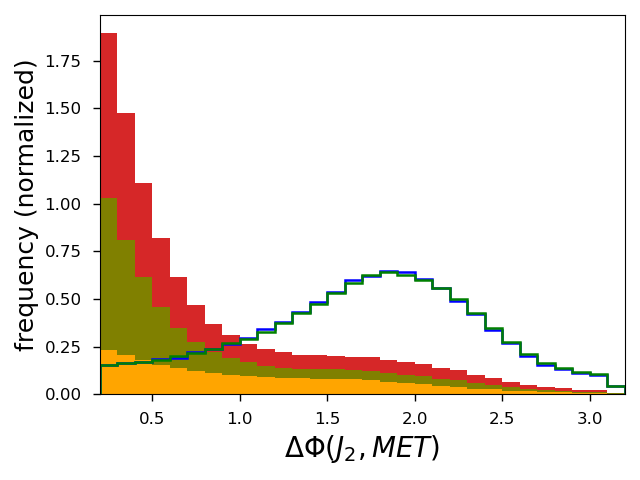}}
	\subfloat[] {\label{DRAK08:DRJ1J2} \includegraphics[width=0.31\textwidth]{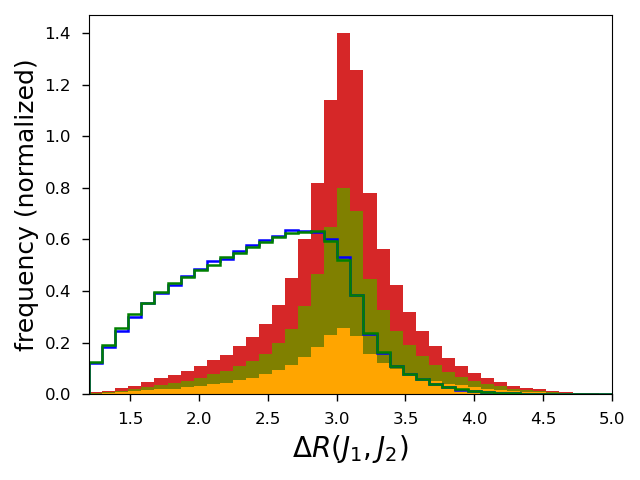}}\\
	\subfloat[] {\label{DRAK08:tau32_J1} \includegraphics[width=0.31\textwidth]{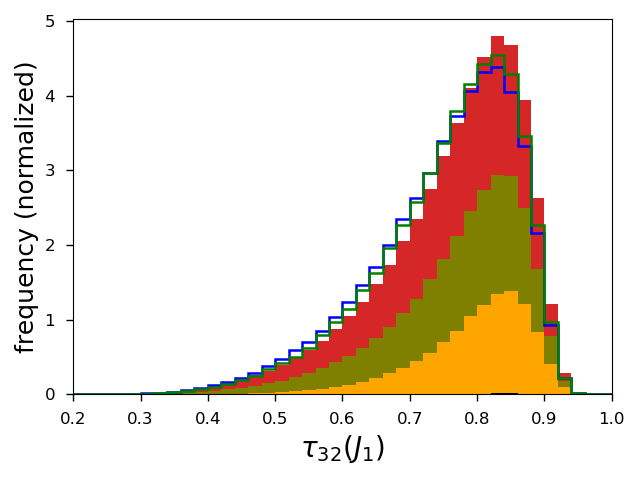}} 
		\subfloat[] {\label{DRAK08:tau32_J2} \includegraphics[width=0.31\textwidth]{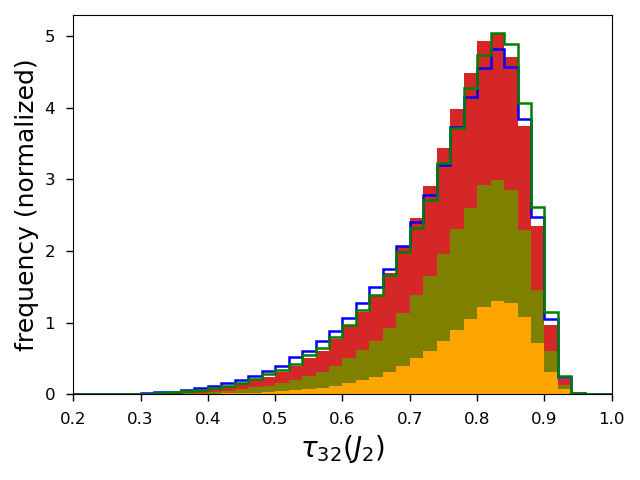}} 
	\subfloat[] {\label{DRAK08:Shat} \includegraphics[width=0.31\textwidth]{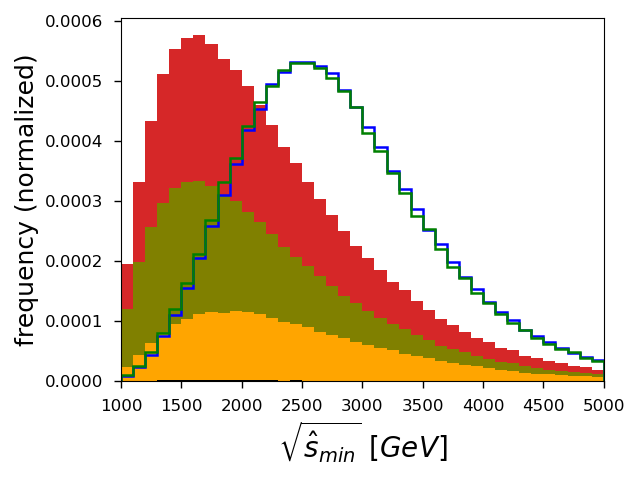}}\\
	\subfloat[] {\label{DRAK08:MT2} \includegraphics[width=0.31\textwidth]{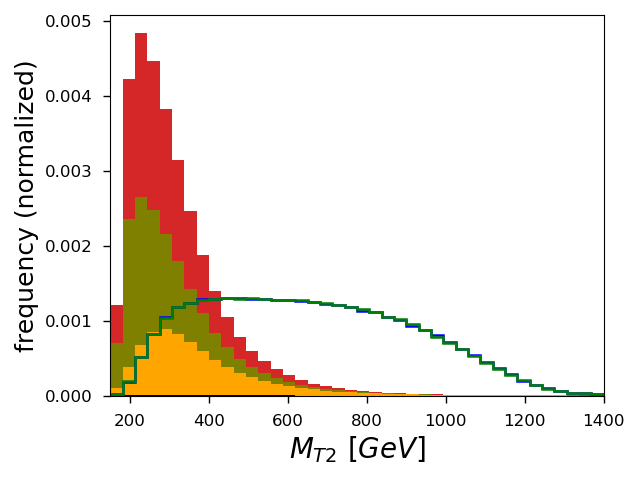}}
	\subfloat[] {\label{DRAK08:pTb} \includegraphics[width=0.31\textwidth]{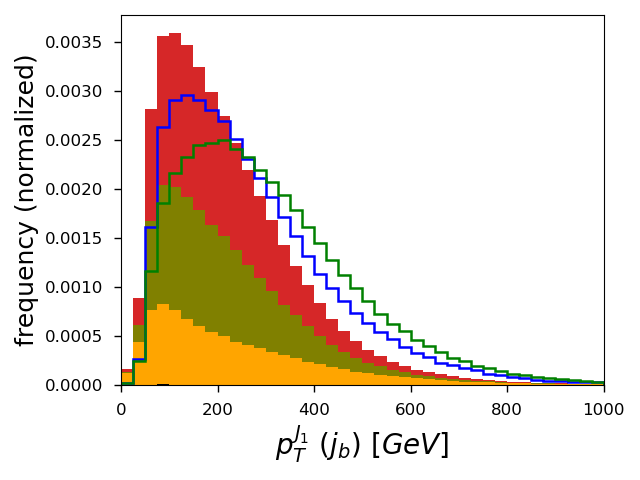}}
	\subfloat[] {\label{DRAK08:InvM_nonb_jets} \includegraphics[width=0.31\textwidth]{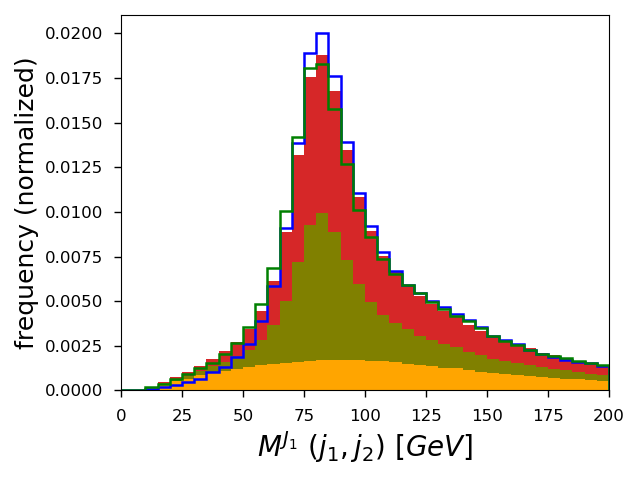}} 
	\caption{Same as Figure \ref{AK12:sig_BG} but for {\sc DRAK08} jets.}
	\label{DRAK08:sig_BG}
\end{figure}

The distributions of the event variables for the dynamic radius jet algorithm are shown in Figure \ref{DRAK08:sig_BG}.

\clearpage
\section{BDT Performance for Fixed Radius Algorithm}
\label{app:BDT}
\begin{figure}[!h]
	\begin{center}
		\includegraphics[width=\textwidth]{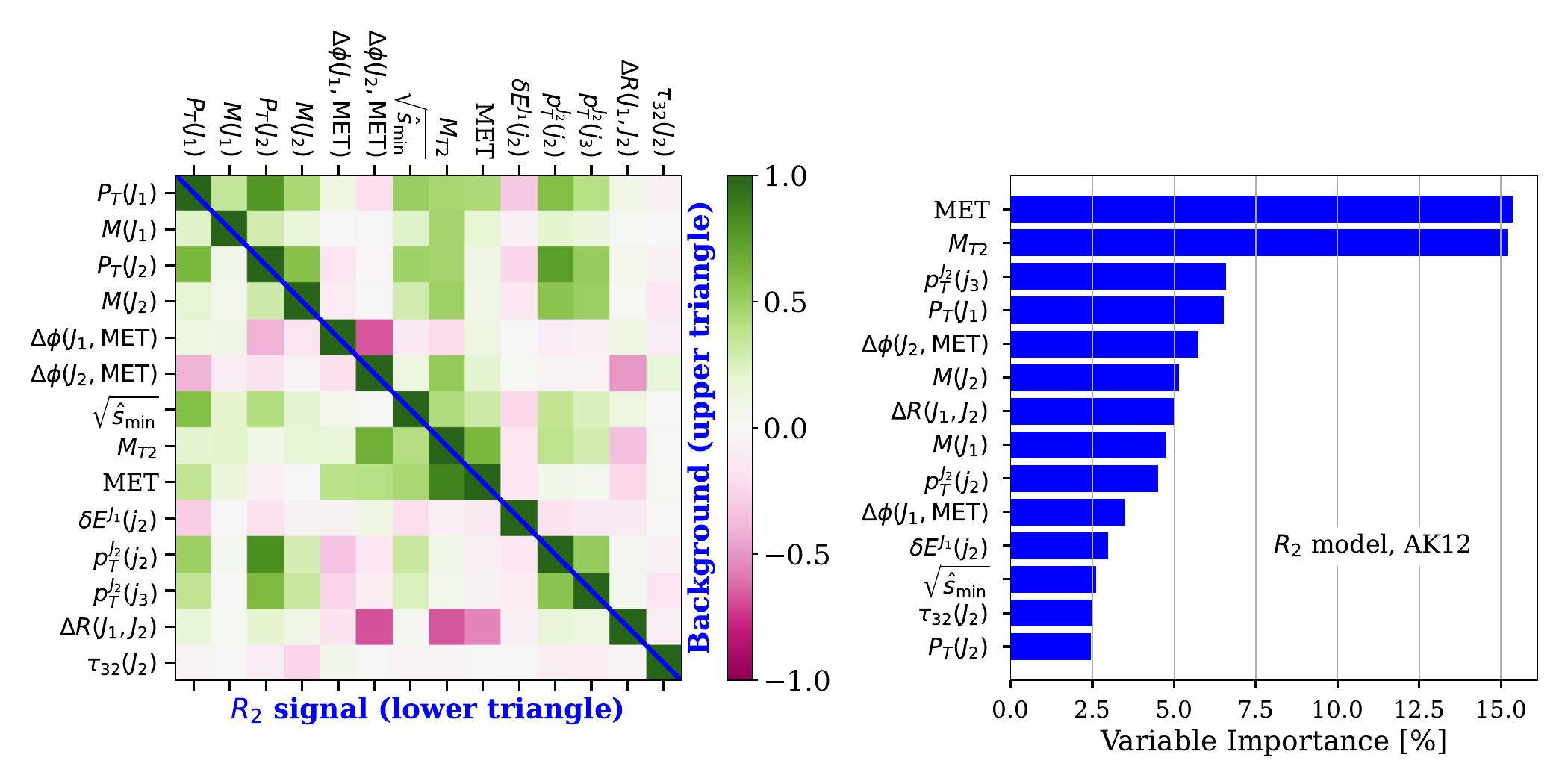}
	\end{center}
\caption{Same as Figure~\ref{R2_DRAK08}, but for the AK12 jet algorithm.}
	\label{R2_AK12}
\end{figure}

The correlation matrix between the variables for the $R_2$ signal and SM backgrounds, along with the variable importances from the BDT training for signal-background separation, are shown in Figure \ref{R2_AK12} for the AK12 jet algorithm.

\begin{figure}[!h]
\begin{center}
	\includegraphics[width=\textwidth]{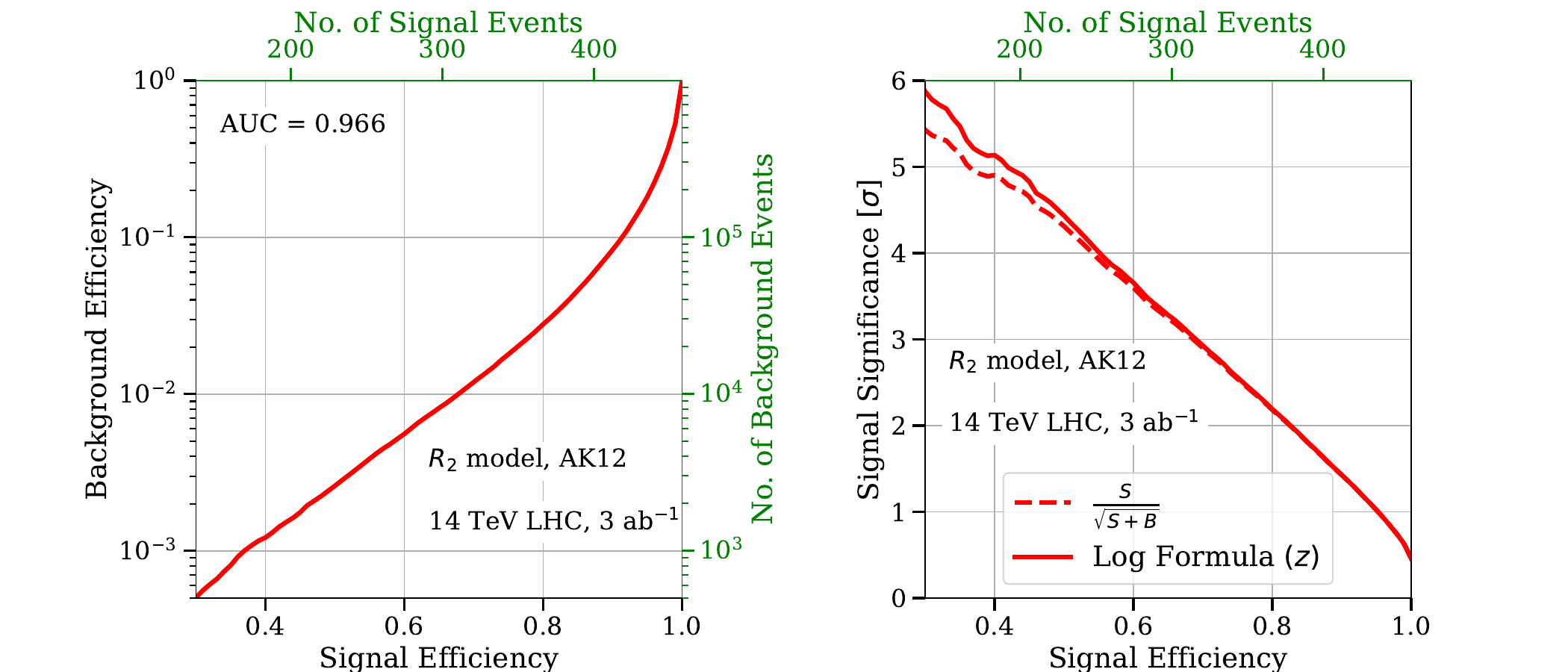}
\end{center}
\caption{Same as Figure \ref{DRAK_R2_ROCSig}, but for the AK12 jet algorithm.}
\label{AK_R2_ROCSig}
\end{figure}

The ROC curve for the $R_2$ signal versus background separation, along with the signal significance as a function of signal efficiency, are shown in Figure \ref{AK_R2_ROCSig} for the AK12 jet algorithm.

\begin{figure}[!h]
	\begin{center}
		\includegraphics[width=\textwidth]{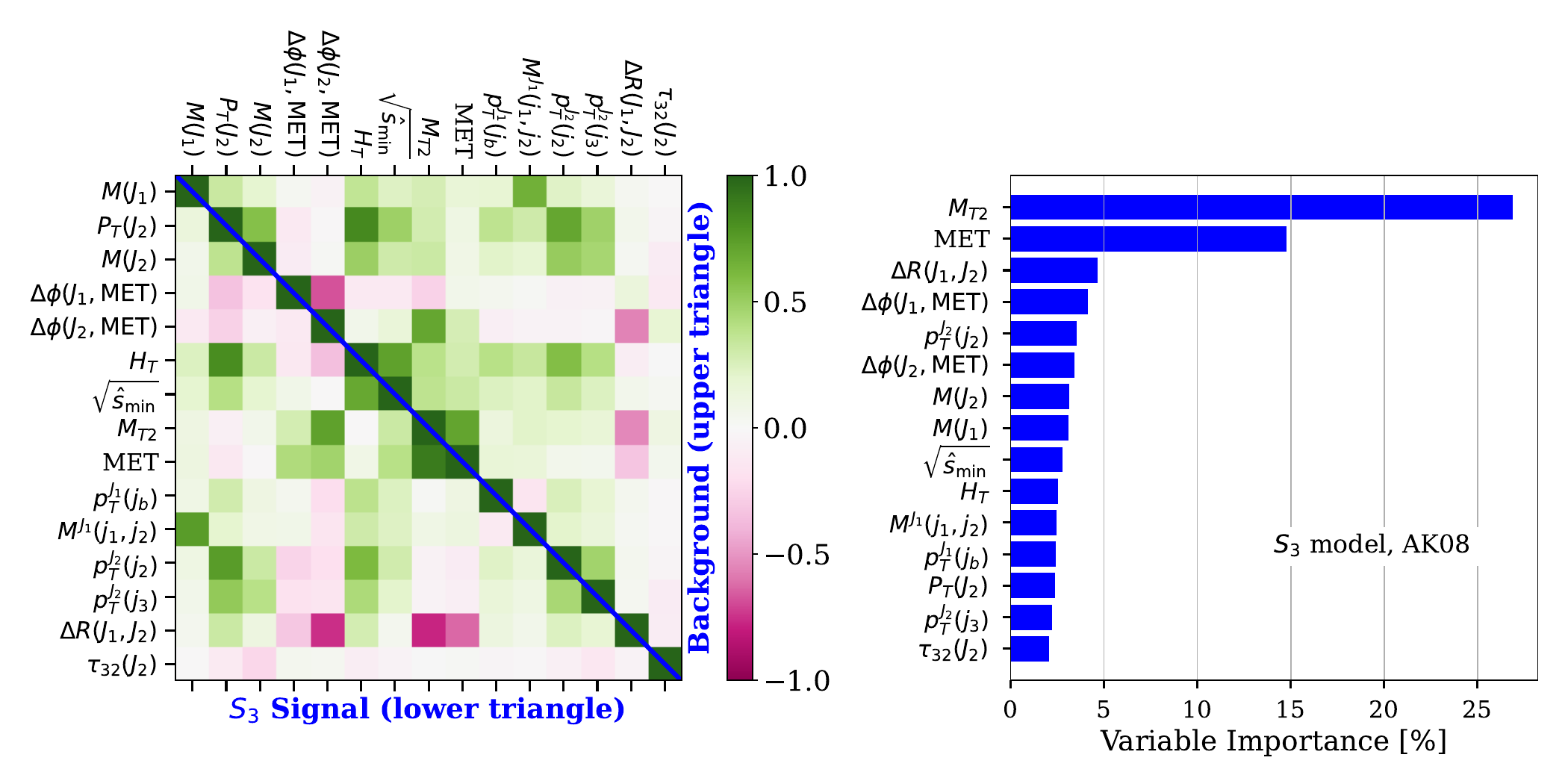}
	\end{center}
\caption{Same as Figure \ref{S3_DRAK06}, but for the AK08 jet algorithm.}
	\label{S3_AK08} 
\end{figure}

The correlation matrix between the variables for the $S_3$ signal and SM backgrounds, along with the variable importances from the BDT training for signal-background separation, are shown in Figure \ref{S3_AK08} for the AK08 jet algorithm.

\begin{figure}[!h]
\begin{center}
	\includegraphics[width=\textwidth]{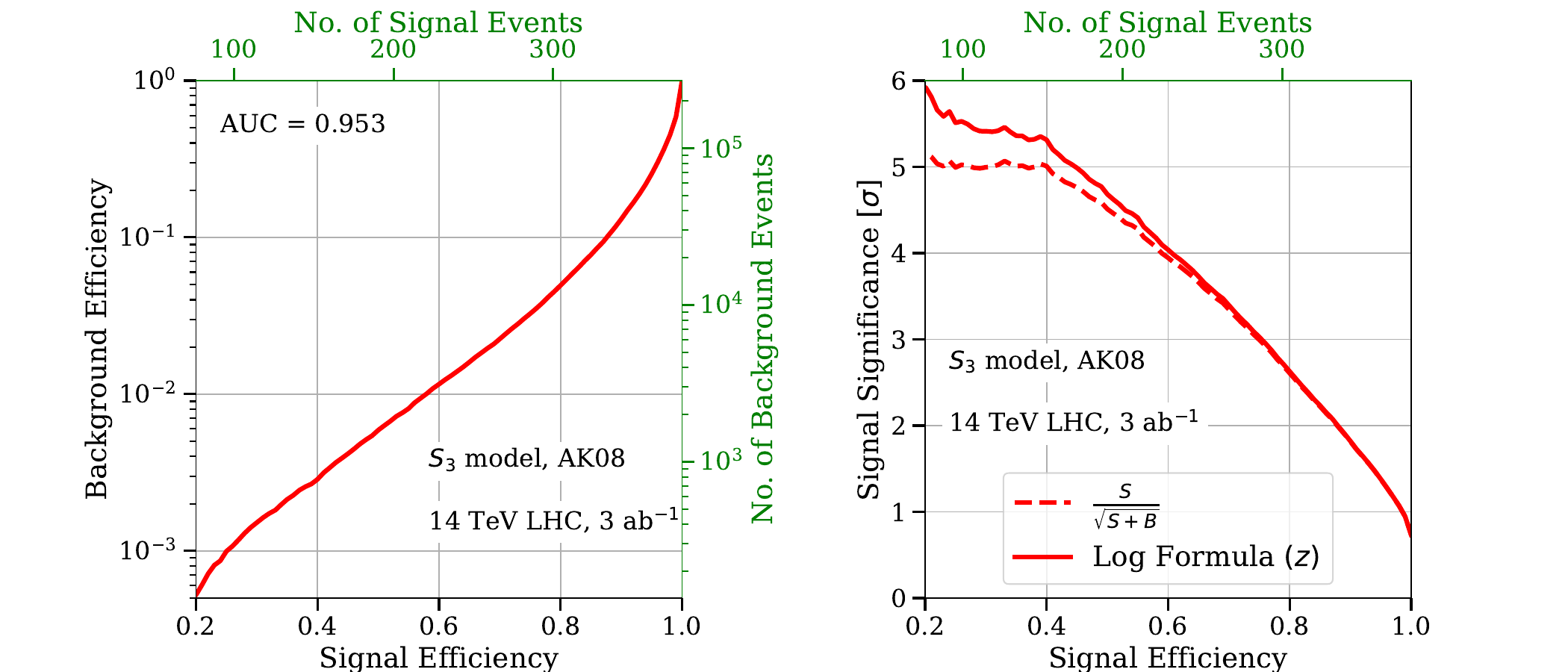}
\end{center}
\caption{Same as Figure \ref{DRAK_S3_ROCSig}, but for the AK08 jet algorithm.}
\label{AK_S3_ROCSig}
\end{figure}

The ROC curve for the $S_3$ signal versus background separation, along with the signal significance as a function of signal efficiency, are shown in Figure \ref{AK_S3_ROCSig} for the AK08 jet algorithm.

\clearpage
\section{Polarization Variables for Fixed Radius Algorithm}
\label{app:polvars}

\begin{figure}[!h]
	\centering
	\includegraphics[width=0.96\textwidth]{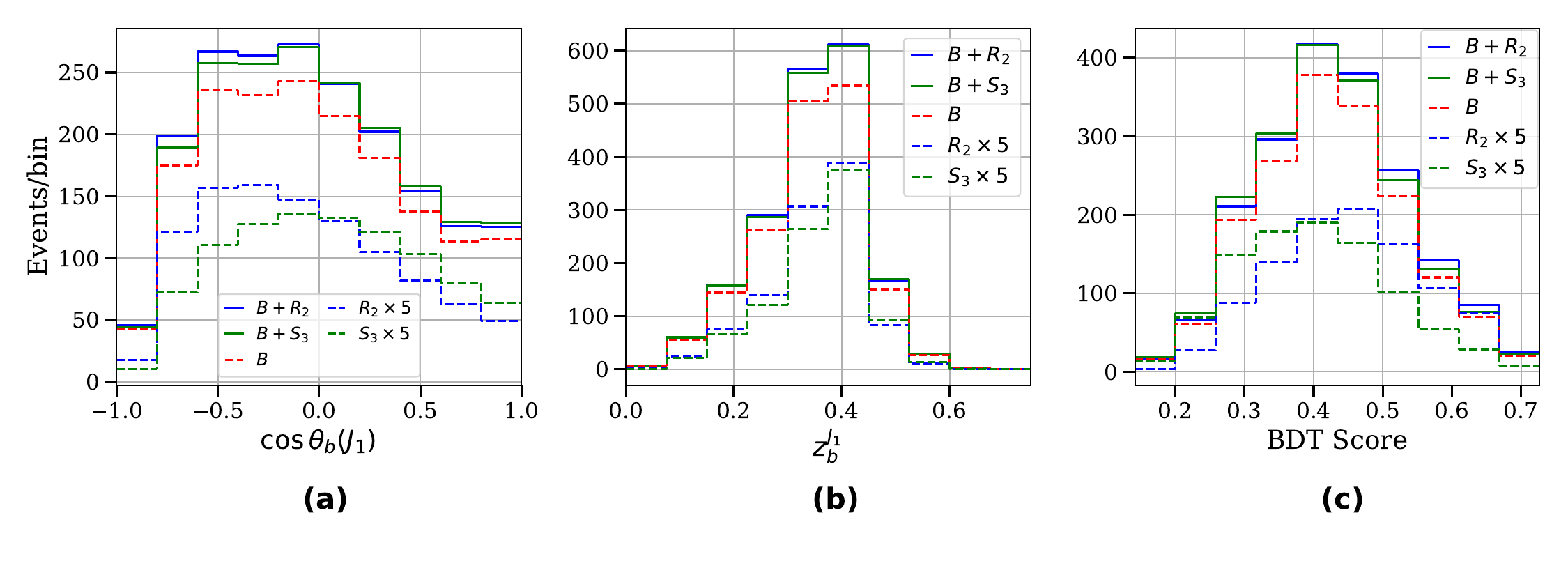}
	\caption{Same as Figure \ref{fig:excluR2DRAK08}, but for the AK12 jet algorithm.}
	\label{fig:excluR2AK12}
\end{figure}

Distributions of polarization-sensitive variables in the signal region are shown in Figure~\ref{fig:excluR2AK12} for the AK12 jet algorithm.

\begin{figure}[!h]
\centering
\includegraphics[width=0.96\textwidth]{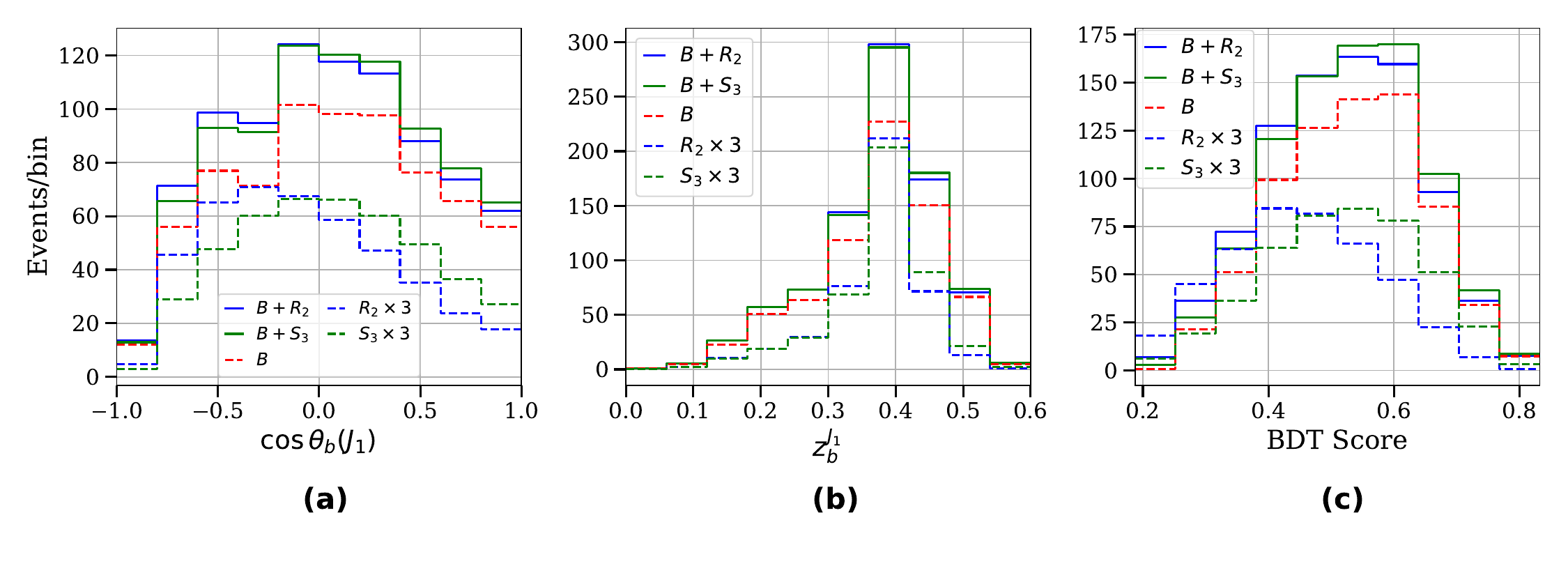}
\caption{Same as Figure \ref{fig:excluS3DRAK06}, but for the AK08 jet algorithm.}
\label{fig:excluS3AK08}
\end{figure}

Distributions of polarization-sensitive variables in the signal region are shown in Figure \ref{fig:excluS3AK08} for the AK08 jet algorithm.

\bibliographystyle{JHEP}
\bibliography{Reference.bib}
\end{document}